\documentclass[a4paper,11pt]{article}

\usepackage{jcappub} 

\usepackage[T1]{fontenc} 
\usepackage{jabbrv}

\title{Strongly Coupled Dark Energy with Warm dark~matter vs.~LCDM}

\author[a]{S. A. Bonometto,}
\author[a]{M. Mezzetti}
\author[b]{and R. Mainini}

\affiliation[a]{INAF, Osservatorio di
  Trieste \& Trieste University, Physics Department, Astronomy Unit, 
\\ Via Tiepolo 11, 34143 Trieste, Italy}
\affiliation[b]{Physics Department G.~Occhialini,
  Milano--Bicocca University,
\\ Piazza della Scienza 3, 20126 Milano, Italy}

\emailAdd{bonometto@oats.inaf.it}
\emailAdd{mezzetti@oats.inaf.it}
\emailAdd{roberto.mainini@mib.infn.it}

\abstract{Cosmologies including strongly Coupled (SC) Dark Energy (DE)
  and Warm dark matter (SCDEW) are based on a conformally invariant
  (CI) attractor solution modifying the early radiative expansion.
  Then, aside of radiation, a kinetic field $\Phi$ and a DM component
  account for a stationary fraction, $\sim 1\, \%$, of the total
  energy.
  Most SCDEW predictions are hardly distinguishable from LCDM, while
  SCDEW alleviates quite a few LCDM conceptual problems, as well as
  its difficulties to meet data below the average galaxy scale.
  The CI expansion begins at the end of inflation, when $\Phi$ (future
  DE) possibly plays a role in reheating, and ends at the Higgs
  scale. Afterwards, a number of viable options is open, allowing for
  the transition from the CI expansion to the present Universe. In
  this paper: (i) We show how the attractor is recovered when the spin
  degrees of freedom decreases. (ii) We perform a detailed comparison
  of CMB anisotropy and polarization spectra for SCDEW and LCDM,
  including tensor components, finding negligible discrepancies. (iii)
  Linear spectra exhibit a greater parameter dependence at large
  $k$'s, but are still consistent with data for suitable parameter
  choices. (iv) We also compare previous simulation results with fresh
  data on galaxy concentration. Finally, (v) we outline numerical
  difficulties at high $k$. This motivates a second related paper \cite{BM2},
  where such problems are treated in a quantitative way.}

\begin{document}
\maketitle
\flushbottom

\section{Introduction}
LCDM models account for a large deal of data, well beyond cosmic
acceleration and background component distribution. In turn, they have
fine tuning and coincidence problems. Accordingly, it is reasonable to
take them as a sort of highly performing {\it effective} models.

It is possible that the true physics laying behind LCDM has to do with
limits to General Relativity (GR) validity \cite{GRV}, namely at large
scales or low densities. Reality could be even more awkward, involving
further dimensions or colliding branes \cite{GRM}.

Here we discuss a less intriguing option, with a feature in common
with several previous suggestions \cite{DEP,amendola}, i.e. that Dark
Energy (DE) is a scalar field $\Phi$, so exploiting a specific feature
of these fields when self--interacting: a negative pressure
approaching their energy density ($|p_\Phi| \sim |\rho_\Phi|$).

Apart of that, our option, dubbed SCDEW (Strongly Coupled Dark Energy
plus Warm dark matter), exhibits key differences from former models:
(i) Starting from the finding of an early conformally invariant (CI)
attractor, SCDEW includes a stationary contribution $\cal O$$(1\, \%)$
of $\Phi$ and Dark Matter (DM), {\it all through radiative eras}; in
fact, if uninteracting, cold DM and a kinetic scalar field $\Phi$
would dilute as $a^{-3}$ and $a^{-6}$, respectively; SCDEW allows for
a suitable energy flow from DM to $\Phi$, so that both components
dilute as $a^{-4}$, in parallel with radiation, along the
attractor. (ii) $\Phi$, besides of yielding DE, has a role in
post--inflationary reheating; this is a new point illustrated in this
work; the option of $\Phi$ being itself the inflaton is not discussed
here. (iii) Today, $\Phi$ must be self--interacting, but the shape of
the self--interaction potential $V(\Phi)$ is unessential and its
(hard) determination is not a test for SCDEW models. (iv) SCDEW
includes two Dark Matter (DM) components, that we dub coupled and
warm (coDM and WDM, respectively). The coupling with $\Phi$ of coDM is
{\it initially} quite strong, as will be detailed shortly.

Moreover, WDM and coDM naturally share similar primeval densities and, in
this paper, we shall debate only cases when they have an equal mass
$\cal O$$(100\, $eV).

Some previous papers \cite{BSLV,BM,BMM} were already devoted to SCDEW
cosmologies; n--body simulations based on them were also performed
\cite{MMPB}, finding substantial improvements in the description of
scales below the average galactic scale, in respect to LCDM (a new
comparison with recent galactic concentration data is also included in
this paper). Apart from resuming the basic features of SCDEW models,
this paper is focused on two points: the possible role of $\Phi$ in
post--inflationary reheating and a model upgrade allowing a low--$z$
fading of coupling.

A number of side results will be also presented. In particular, we
provide detailed information on how model parameters influence matter
fluctuation spectra $P(k,z)$ at low redshift $z$. Such parameters are
essentially 2, accounting for $\Phi$--DM coupling and WDM particle
mass. Recent data (e.g., high--$z$ Lyman--$\alpha$ clouds
\cite{pallosobra}, or the Hubble telescope and Planck $H_0$ values
\cite{Xia}) put constraints on them, in a way we shall indicate. A
stringent quantitative analysis is however postponed to further work.

The possibility of a common origin for DM components is also debated
here. This is related to the CI attractor behavior when the fall of
cosmic temperature yields a decrease in the number of effective spin
degrees of freedom
\begin{equation}
g = {7 \over 8} {\cal N}_{fermion} +  {\cal N}_{boson} ~.
\label{g}
\end{equation}
Here ${\cal N}_{fermion,boson}$ is the number of fermion, boson
relativistic spin states. The problem will be illustrated here by
showing the attractor behavior at the Quark--Hadron transition,
occurring at a temperature $T_c \sim 170\, $MeV, when $g$ decreases by
$\sim 61.75$ to $\sim 14.25$ (with 3 neutrino species) as, then, the
contribute of strongly interacting particles falls to almost nil (pion
density is strongly suppressed by their physical size, while baryon
masses are $\gg T_c$).

Another point of this work is a detailed comparison of SCDEW and LCDM
anisotropies and polarization for CMB, also in presence of a tensor
signal.

We shall finally motivate a second related paper, where the behavior
of early non--linearities in coDM will be debated, finding that, far
from yielding early bound systems, a fear outlined in \cite{BMM}, they
are expected to cause fluctuation dissipation, so yielding a
low--scale spectral cutoff, around 10--100$\, M_\odot$. It may be
worth comparing this cutoff scale with the expected spectral cutoff in
ordinary LWDM models, if including DM particles of mass $\sim 100\,
$eV: in such case the $P(k,z)$ cutoff lays well above large galaxy
mass scales.

The plan of the paper is as follows: In the next Section we shall
first resume the SCDEW cosmology scheme, then outlining the mechanism
allowing to exit the CI expansion regime though the Higgs mass
acquisition by the coupled spinor field. This is the basic new idea we
develop in this work and the similarity between the expected Higgs
mass scale of the warm DM component and such mass, suggested us to
debate, in Section 3, the possibility that the 2 fields are somehow
directly related. In the same Section we also show how the kinetic
attractor is recovered when the effective number of relativistic spin
degrees of freedom varies. Section 4 is then devoted to discussing the
linear fluctuation evolution in this version of SCDEW cosmologies. We
therefore show in detail the (small) discrepancies between LCDM and
SCDEW in CMB spectra, as well as the more significant discrecancies in
the linear fluctuation spectra, however showing that this model
cathegory requires a treatment more complicated than a linear program
extension, to treat the high--$k$ expectations. We conclude the paper
with a discussion section.

\section{Coupled DE in the early Universe. Background components}

\subsection{A kinetic tracker solution}
The state equation of a purely kinetic scalar field $\Phi$, whose {\it
  free} Lagrangian
\begin{equation}
{\cal L}_f \sim \partial^\mu \Phi  \partial_\mu \Phi ~,
\label{freelagrangian}
\end{equation}
is
$w=p_k/\rho_k = 1$
($p_k,~\rho_k:$ field kinetic pressure, energy density). Accordingly,
$\rho_{k} = \dot \Phi^2/2a^2 \propto a^{-6}$. Here $a$ is
the scale factor appearing in the background metric
\begin{equation}
ds^2 = a^2(\tau) (d\tau^2 - d\lambda^2)~,
\label{backgroundmetric}
\end{equation}
$\tau$ being the {\it conformal time} and $\lambda$ being the {\it
  comoving distance}. Dots indicate differentiation with respect to
$\tau$, all through the paper.

It is also known that non--relativistic DM density $\rho_c \propto
a^{-3}$, its state parameter being $w=0$. Accordingly, it appears
rather intuitive that a suitable energy flow from DM to $\Phi$ could
speed up DM dilution while $\rho_k$ dilution slows down, so that both
dilute $\propto a^{-4}$. If the overall expansion is radiation
dominated (RD), this yields constant early density parameters
$\Omega_c$ and $\Omega_d$ (for DM and $\Phi$, respectively). The shift
required, for the dilution exponent of the kinetic energy of the field
$\Phi$, is double in respect to DM exponent shift. Therefore, also
intuitively, we expect such effect to require that $\Omega_c = 2\,
\Omega_d$.

As a matter of fact, in \cite{BSLV} it was shown that these
expectations are consistent with a DM--$\Phi$ coupling ruled by the
equations
\begin{equation}
T^{(d)~\mu}_{~~~\, ~~\nu;\mu} = +C T^{(c)} \Phi_{,\nu}~,
~~~~~~~~~~
T^{(c)~\mu}_{~~~~\nu;\mu} =- C T^{(c)} \Phi_{,\nu}~,
\label{relativisticeq}
\end{equation}
an option introduced since the early papers on DE \cite{amendola},
aimed there to ease DE fine tuning and coincidence problems {\it in
  the present epoch,} i.e.  after radiation dominated expansion end.
This option gave origin to a vaste literature and simulations of
these models were also compared with data \cite{macciobaldi}. These
equations will be however used here in a fully different context.

Let us then outline that, in eq.~(\ref{relativisticeq})
$T^{(c,d)}_{\mu\nu}$ are the stress--energy tensors of the DM and
$\Phi$ field components, $T^{(c)}$ is the trace of the former tensor
while
\begin{equation}
C = b/m_p= (16 \pi/3)^{1/2} \beta/m_p~
\label{couplingconstant}
\end{equation}
($m_p:$ the Planck mass) is the DM--DE coupling constant.

As a matter of fact, these very equations, used now in the early {\it
  radiative expansion}, allowed \cite{BSLV} to show that, if the early
density parameters are
\begin{equation}
  \Omega_c = 2\, \Omega_d = {1 \over 2 \beta^2}~,
\label{omegas}
\end{equation}
we are on an attractor: if we start from DM and $\Phi$ components
which exhibit initial densities not fultilling eq.~(\ref{omegas}),
their densities rapidly modify finally settling on such {\it
  conformally invariant} solution which is, therefore, a {\it kinetic
  tracker solution} in the radiation dominated epoch. 

In Section 3 we shall again check the attractor efficiency, by
reporting the evolution of coupled DM and DE while the cosmological
Quark--Hadron transition led to a sudden decrease of the spin degrees
of freedom in the radiative component. Here we shall partially review
results obtained in a previous paper \cite{universe}, by treating,
however, different parameter ranges.

In the frame of reference where the metric is
(\ref{backgroundmetric}), eqs.~(\ref{relativisticeq}) read
\begin{equation}
\ddot \Phi + 2{\dot a \over a} \dot \Phi = -a^2 V'_\Phi + C a^2 \rho_c
~,~~~~~~~ \dot \rho_c + 3 {\dot a \over a} \rho_c = -C \rho_c \dot \Phi~,
\label{nonrelativisticeq}
\end{equation}
$V(\Phi) $ being a self--interaction potential density for the $\Phi$
field which, however, is nil or negligible along the kinetic tracker
solution.

These equations, with selected $V(\Phi)$ potentials, allow for
high--$z$ tracker solutions. They belong to the approach cited above
\cite{amendola,macciobaldi}. The attractor considered herebelow is for
a purely kinetic $\Phi$ field, so that the potential shape is fully
irrelevant.

\subsection{Coupled cosmologies with fixed $w(a)$}
The quest for an expression of the $V(\Phi)$ potential, in this
context, might comes much after. Present and future observations
\cite{SNIA,demianski,lensing}, however, more directly constrain the
state parameter $w(a)$, rather than $V(\Phi)$. Accordingly, in
\cite{BSLV}, it was shown that the eqs.~(\ref{nonrelativisticeq}) also
read
\begin{equation}
  \dot \Phi_1 + \tilde w{\dot a \over a} \Phi_1 = {1+w \over 2} C a^2 \rho_c
  ~,~~~~~~~ \dot \rho_c + 3 {\dot a \over a}\rho_c = -C \rho_c \Phi_1~,
\label{nonrelativisticeq1}
\end{equation}
while
\begin{equation}
2 \tilde w = 1+3w - {d \ln(1+w) \over d\ln a}
\label{wtilde}
\end{equation}
and $\Phi_1 \equiv \dot \Phi$. These equations aim to tell us $\Phi$
evolution directly from $w(a)$ dependence. Notice that they do not
involve the undifferentiated field $\Phi$, so that the whole system is
just second order, and $w(a)$ information always admits an arbitrary
additional constant~on~$\Phi$.

This approach simplifies our analysis also for the very--early
Universe context, when we assume $w \equiv 1$ (purely kinetic field)
and a radiative expansion. At a suitable low redshift, however, there
must be a transition from kinetic to potential dominance, so that the
$\Phi$ can account for a component with $w \simeq -1$ today.

This is not an {\it ad--hoc} feature of SCDEW. Also coupled DE
approaches based on specific potentials (e.g., a SUGRA or RP) require
a fast transition of $\Phi$ from potential to kinetic regime above a
suitable $z_{kp}$, unless a very low $\beta$ ($\lesssim 10^{-3}$) is
selected. Examples were shown in previous work, so that requiring
\begin{equation}
w(a) = {1 - A \over 1 + A} ~~~{\rm with} ~~~ A = \left(a \over a_{kp}
\right)^\epsilon~,
\label{kp}
\end{equation}
as suggested in \cite{BM}, will hardly contradict any data in the next
decades; here $a_{kp}=1/(1+z_{kp})$. Selecting $\epsilon$ is then as
choosing a potential expression. The very dependence on $\epsilon$ of
results is however extremely mild (see again \cite{BM}) and all
through this paper we take $\epsilon = 2.9$~.

\subsection{Lagrangian interactions and inflationary reheating}
Once we assume $w \equiv 1$ during the radiative expansion,
eqs.~(\ref{nonrelativisticeq1}) can be soon integrated, yielding the
above mentioned kinetic tracker solution
\begin{equation}
\Phi_1 = 1/C\tau~,~~~~\rho_c = \bar \rho_{c}(\bar a/a)^4~,
\label{solution}
\end{equation}
wherefrom we deduce that 
\begin{equation}
C\Phi = \ln(\tau/\bar \tau)
\label{logarithmicphi}
\end{equation}
$\Phi$ being again undetermined for an arbitrary additive constant; in
other terms, we ignore the field value at $\bar \tau$; $\bar \rho_c$
and $\bar a$ are the energy density and the scale factor~at~$\bar
\tau$.

In \cite{BMM} it was then shown that the coupling term in
eqs.~(\ref{nonrelativisticeq}) or (\ref{nonrelativisticeq1}) derives
from a generalized Yukawa interaction Lagrangian
\begin{equation}
{\cal L}_m = - \mu f(\Phi/m) \bar \psi \psi
\label{interaction}
\end{equation}
provided that
\begin{equation}
f = \exp(-\Phi/m)~.
\label{fff1}
\end{equation}
Here 2 independent mass scales, $m = m_p/b$ and $\mu = g\, m_p$ need
to be introduced for dimensional reasons. The constant $b$ coincides
with the factor $b$ gauging the DM--$\Phi$ interaction strength in
eq.~(\ref{couplingconstant}), so that $C=b/m$; on the contrary, $g$
keeps undetermined as well as a $\Phi$ additive constant. Here, the
spinor field $\psi$ accounts for interacting DM, its particle number
density operator being $n \propto \bar \psi \psi$. From this
Lagrangian density we work out the energy density
\begin{equation}
\rho_c  = \mu f(C\Phi) \bar \psi \psi~,
\end{equation}
(formally $= -{\cal L}_m$) for DM spinor quanta, which are
non--relativistic. It is then worth focusing on the term
\begin{equation} 
  {\delta {\cal L}_m \over \delta \Phi} \equiv [{\cal L}_m]'_\Phi = -
  \mu {f'}_\Phi (C\Phi) \bar \psi \psi = - {{f'}_\Phi (C\Phi) \over
    f(C\Phi)} \rho_c = C\rho_c
\label{eulerolagrange}
\end{equation}
of the Euler--Lagrange equation which, multiplied by $a^2$, is the
second term at the r.h.s. of the first eq.~(\ref{nonrelativisticeq})
(let us remind that $C=b/m_p$). This will become a critical point when
we shall consider the effects of Higgs mass acquisition by the
different fields.

Meanwhile, let us notice that
\begin{equation}
\rho_c \propto f(C\Phi) a^{-3},
\label{rhosc}
\end{equation}
according also to the findings in \cite{das2006},
as indeed $n
\propto a^{-3}$. Taking into account the expression
(\ref{logarithmicphi}) of $\Phi$, we then confirm that
\begin{equation}
\rho_c a^4 = \bar \rho_c \bar a^4 
\label{a^4}
\end{equation}
during the radiative era.

Let us now consider the coupling (\ref{interaction}) within the
context of large--field inflation. We can imagine that, around Planck
time, the ``initial'' $\Phi = F \times m_p$ with $F = 3$--4~. From the
expressions (\ref{interaction}) and (\ref{fff1}), we see that the
interaction Lagrangian, initially reading
\begin{equation}
{\cal L}_m \simeq - \mu \exp(-Fb) \bar \psi \psi~,
\label{LI}
\end{equation}
is then strongly suppressed (namely if $b$ values are large, yielding
a later very strong coupling) and the interaction acquires strength
only when $\Phi$ rolls down to values $\sim m_p$ or less. Let us state
that elsehow: The Lagrangian (\ref{LI}) yields two equilibrium
configurations: $\Phi \to \infty$ and $\Phi$ on the CI solution.
During inflation, the former equilibrium is approached; but $\Phi$
decrease drives away from it, finally favoring reheating; $\Phi$
decrease, however, does not proceed to nil: it stops when the latter
configuration is attained.

Accordingly, the Lagrangian (\ref{LI}) should not account for the slow
rolling down process itself, just in the same way as, later on, it
shall not account for DE energy density. On the contrary, it accounts
for fastly turning the $\Phi$ field energy into DM, i.e. into $\psi$
quanta, as it occurs during the reheating; then, later on, it accounts
for the same process at a suitably tuned slow rate, in the stationary
expansion regime. Until such a regime is reached, this very process
will receive further quantum contributions from the symultaneous
evolution of the vacuum state, accounted for by Bogoliubov
transformations, a process also stopping at the reach of the
stationary expansion regime along the attractor, in a conformally
invariant (CI) evolution.

The potential term driving the inflationary process could rather (at
least partially) coincide with the one yielding DE. This however
requires a suitable evolution of the constant(s) it contains, so that
the passage of $\Phi$ from potential to kinetic dominance (and
viceversa) occurs at different $\Phi$ values.  Owing to the
logarithmic evolution of $\Phi$ (\ref{logarithmicphi}), however, the
constant(s) are just required to evolve logarithmically, as can be
consistent with renormalization group expectations.

A toy model possibly accounting for these behaviors is being analysed
and will be discussed in a forthcoming paper.

\subsection{DM field components, Higgs mass acquisition, and coupling
  late fading} 
The relation between the Lagrangian (\ref{interaction}) and the
post--inflationary reheating was not outlined in previous work.
Rather, we treated such Lagrangian as just yielding a peculiar
$z$--dependent mass term for the DM field $\psi$.

Another DM component, made of low mass $\cal O$$(100\, $eV) quanta (of
a second spinor field $\psi_1$), was then introduced. It was then
noticed that, if $\psi_1$ quanta have a mass in such range, the two DM
components unavoidably share close primeval densities.

As is natural to expect that $\psi_1$ quanta acquired such mass $m_w$
at the electroweak (EW) transition scale, through the Higgs mechanism,
here we shall suppose that also $\psi$ acquires a similar mass at the
same scale. There is no specific reason to suppose that $\psi$ and
$\psi_1$ will share identical mass values, an assumption somehow
envisaging a related nature of the two dark components, but we can
exploit the model degrees of freedom and keep to such an option, to
select model parameters for this work.

{ The point is that $\psi$ acquiring a Higgs mass bears critical
consequences for the later evolution of DM--DE coupling and, through
it, for today's component densities. In fact, at $T<T_{EW}$, the
Lagrangian (\ref{interaction}) becomes}
\begin{equation} 
\label{higgslagrangian}
\tilde {\cal L}_m = -[\mu f(\Phi /m) + \tilde \mu ] \bar \psi \psi
\equiv -\mu [\exp(-C\Phi) + \tilde \mu/\mu] \bar \psi \psi~.
\end{equation}
By using this Lagrangian we can then re--do the functional
differentiation operation in eq.~(\ref{eulerolagrange}), so obtaining
\begin{equation}
{\delta \tilde{\cal L}_m \over \delta \Phi} =-\frac{f'(C\Phi
  )}{f(C\Phi)+{\tilde \mu/\mu}}\rho_c = {C \over 1 + {\cal R}
  \exp[C(\Phi-\bar \Phi)]} \rho_c ~.
\label{variablecoupling}
\end{equation}
Here
\begin{equation}
\label{calR}
{\cal R} = (\tilde \mu/\mu) \exp (C \bar\Phi)~,
\end{equation}
$\bar\Phi$ being the value of the field at a suitable reference time,
e.g. during the stationary regime. However, rather than setting a
value for $\bar \Phi$, it is more convenient to introduce a
coefficient $g_h$ such that
\begin{equation}
\exp(C\bar \Phi) = \mu/(g_h m_p)
\label{gh}
\end{equation}
so yielding
\begin{equation}
{\cal R} = \tilde \mu/(g_h m_p)~.
\label{gh1}
\end{equation}
Let us also outline that, if the reference time is changed from $\bar
\tau_1$ to $\bar \tau_2$, both assumed to belong to the CI expansion
epoch, it shall be
\begin{equation}
{\cal R}_1 = {\cal R}_2 {\bar \tau_1}/{\bar \tau_2}~~~~{\rm and}~~~~
g_{h,1} = g_{h,2} {\bar \tau_2}/{\bar \tau_1}~.
\end{equation}
The key point, however, is that the dynamical equations, even in the
presence of a Higgs' mass for the $\psi$ field, keep the forms
(\ref{nonrelativisticeq}) or (\ref{nonrelativisticeq1}), provided that
we perform the replacements
\begin{equation}
  C \to C_{eff} = {C \over 1 + {\cal R} \exp[C(\Phi-\bar \Phi)]}~~~{\rm
    and/or}~~~ \beta \to \beta_{eff} = {\beta \over 1 + 
 {\cal R} \exp[C(\Phi-\bar \Phi)]}~.
\label{newC}
\end{equation}
Then, as soon as the $\Phi$ increase causes $\Phi-\bar \Phi$ to
approach $-\ln({\cal R})/C$, the denominators in eq.~(\ref{newC})
differ from unity, so suppressing the effective coupling intensity.
This gives a new dynamical significance to the value of $\Phi:$ in the
absence of the Higgs mass $\tilde \mu$, it would be significant only
after kinetic--potential transition, at the eve of the present epoch,
and, possibly, before the potential--kinetic transition, in the late
inflationary expansion.

Numerical results are then obtainable by integrating the system made
by eqs.~(\ref{nonrelativisticeq1}), the Friedmann equation, and the
``trivial'' equation $d (\Phi-\bar\Phi)/d\tau = \Phi_1$, once the
$\beta $ shift (\ref{newC}) is done. 

In order to set initial conditions, we may then ideally extend the
CI regime down to the Planck time $\tau_p$, so setting the initial
field value $\Phi_i = \Phi_p\, \ln(\tau_i/\tau_p)$, while we take $g_h
= 2\pi$. In order to have $\tilde \mu = {\cal O}(100\, $eV), we then
have ${\cal R} \sim 10^{-27}$, a typical value for the ratio between
heavy neutrino and Planck mass scales.

The decline of $\beta_{eff}$
\begin{figure}[t!]
\begin{center}
\vskip -2.5truecm
\includegraphics[height=10.cm,angle=0]{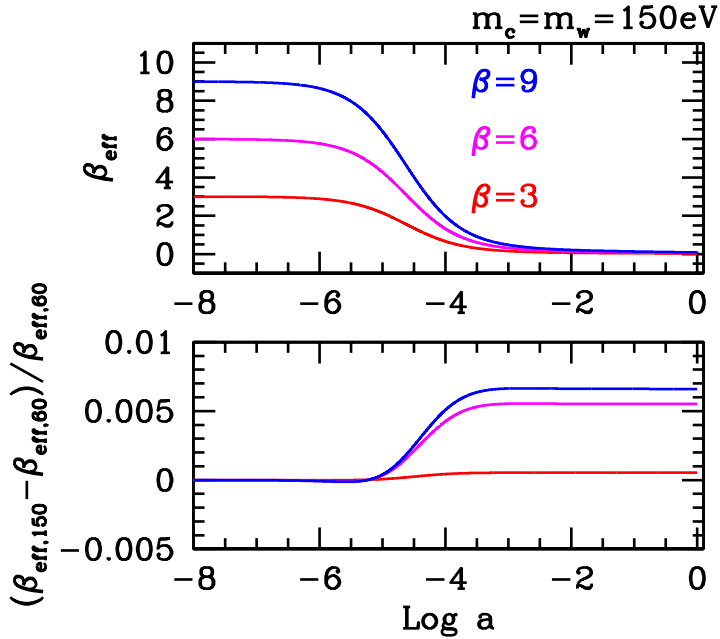}
\end{center}
\vskip -.5truecm
\caption{Scale dependence of the effective coupling in SCDEW
  cosmologies with $m_w = 100\, $eV for various couplings $\beta$. }
\label{betaeff}
\vskip +.1truecm
\end{figure}
\begin{figure}[h!]
\begin{center}
\vskip -1.5truecm
\includegraphics[height=12.5cm,angle=0]{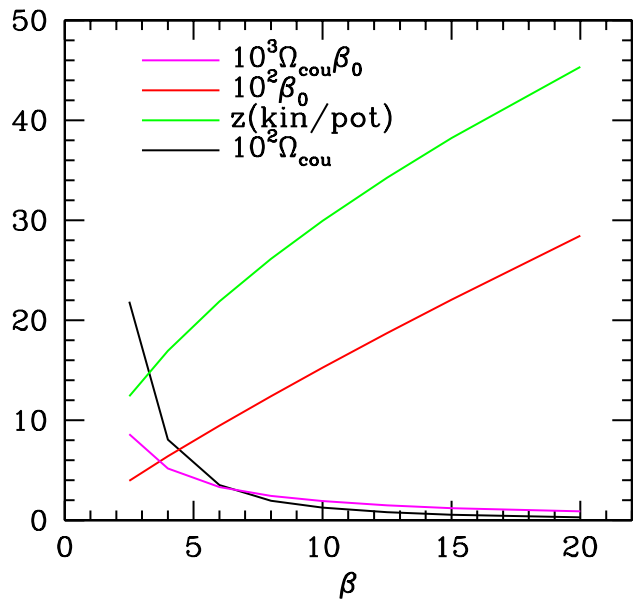}
\end{center}
\vskip -3.1truecm
\caption{For various (high--$z$) couplings $\beta$ and $m_w = 100\,
  $eV, we plot the $z=0$ values of $\beta_0$ and $\Omega_{cou}$. The
  redshift where the kin--pot transition must occur are also
  plotted. }
\label{zkp}
\vskip +.1truecm
\end{figure}
is illustrated in Figure \ref{betaeff}. In Figure \ref{zkp} we then
show the values $\beta_0$ attained by coupling at $z=0$ for $\tilde
\mu = 100\, $eV. For such $\tilde \mu$, even for $\beta = 20$ the
coupling at $z=0$ does not overcome 0.3 (notice that low--$z$ limits
on coupling \cite{Xia} do not apply to this case, as coDM is not the
only DM component).  The Figure also shows the density parameter
$\Omega_{0,cou}$ of coupled DM at $z=0$, so outlining that, starting
from a significant fraction of the total cosmic DM when $\beta \sim
3$, it accounts for $<2$--$3\, \%$ of DM (itself accounting for $\sim
25\, \%$ of cosmic density) for $\beta >\sim 8$. If we consider the
product $\beta_0 \Omega_{0,cou}$, we see that it is also gradually
decreasing. This product gauges the ``average coupling'' of the whole
(warm and cold) DM, whose total density parameter is kept constant in
the models considered.

It is also worth outlining that results shown in this work are
obtained by assuming $\Omega_d=0.7$, $\Omega_b = 0.045$, Hubble
constant $h = 0.685$, $T_0 = 2.726\, $K, optical depth 0.089, $n_s =
0.986$ and 3 (almost massless) neutrino species.

\section{A common origin for DM components?}
Taking ``equal'' Higgs masses for the two DM components (coupled and
uncoupled) alludes to their possible common nature, as though they
would differ just in the strength of the coupling to $\Phi$.  In turn,
this option follows from the observation that viable SCDEW models are
mostly characterized by close densities of the 2 DM components at high
$z$. 

This last effect, however, requires a deeper analysis. Let us suppose,
e.g., that the component uncoupled to $\Phi$ decoupled from all other
cosmic components at a suitable redshift $z_d$. At that time it is
reasonable to expect $\rho_w \simeq \rho_c$ ($\rho_w$, $\rho_c:$
uncoupled and coupled DM densities, respectively). Later on,
\begin{equation}
\rho_w(z) = \rho_w(z_d)\, \, \left[ (1+z)/(1+z_d) \right]^4
\label{rhoz}
\end{equation}
until derelativization. On the contrary, $\rho_c(z) \simeq
\rho_{cr}(z) /2\beta^2$ ($\rho_{cr}(z):$ critical density at $z$).  As
a consequence of entropy conservation, at any decrease of $g$ (defined
in eq. 1.1) after $z_d$, $\rho_{cr}$ has therefore a jump upward
$\propto g^{-1/3}$ and, although after a suitable time delay, $\rho_c$
is bound to accompany such shift, to keep on the attractor.
\begin{figure}[h!]
\begin{center}
\vskip -4.5truecm
\includegraphics[height=10.5cm,angle=0]{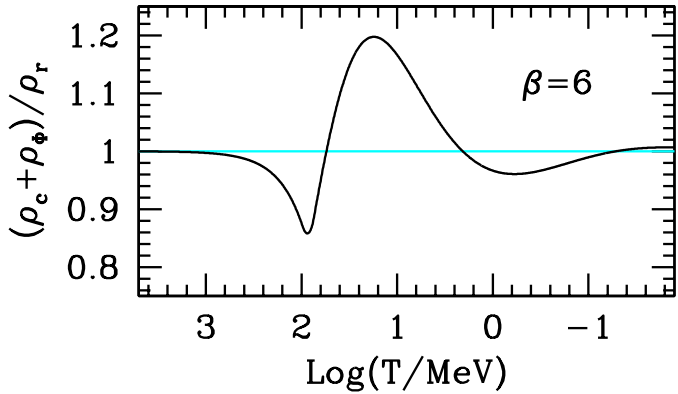}
\end{center}
\vskip -.5truecm
\caption{Time evolution of the ratio between densities of coupled DM
  plus $\Phi$ and radiation, in the proximity of the QH transition,
  for a SCDEW cosmology (with $m_w=100$ and $\beta=6$), normalized to
  $3/4\beta^2 = 0.02083$. The attractor is recovered after a few
  oscillations, which could however affect BBN redshift range. For
  $\beta=6$, effects on BBN are approximately one quarter of what is
  due to electron--positron annihilation. }
\label{rr6}
\vskip -.1truecm
\end{figure}

\begin{figure}[h!]
\begin{center}
\vskip -.5truecm
\includegraphics[height=7.25cm,angle=0]{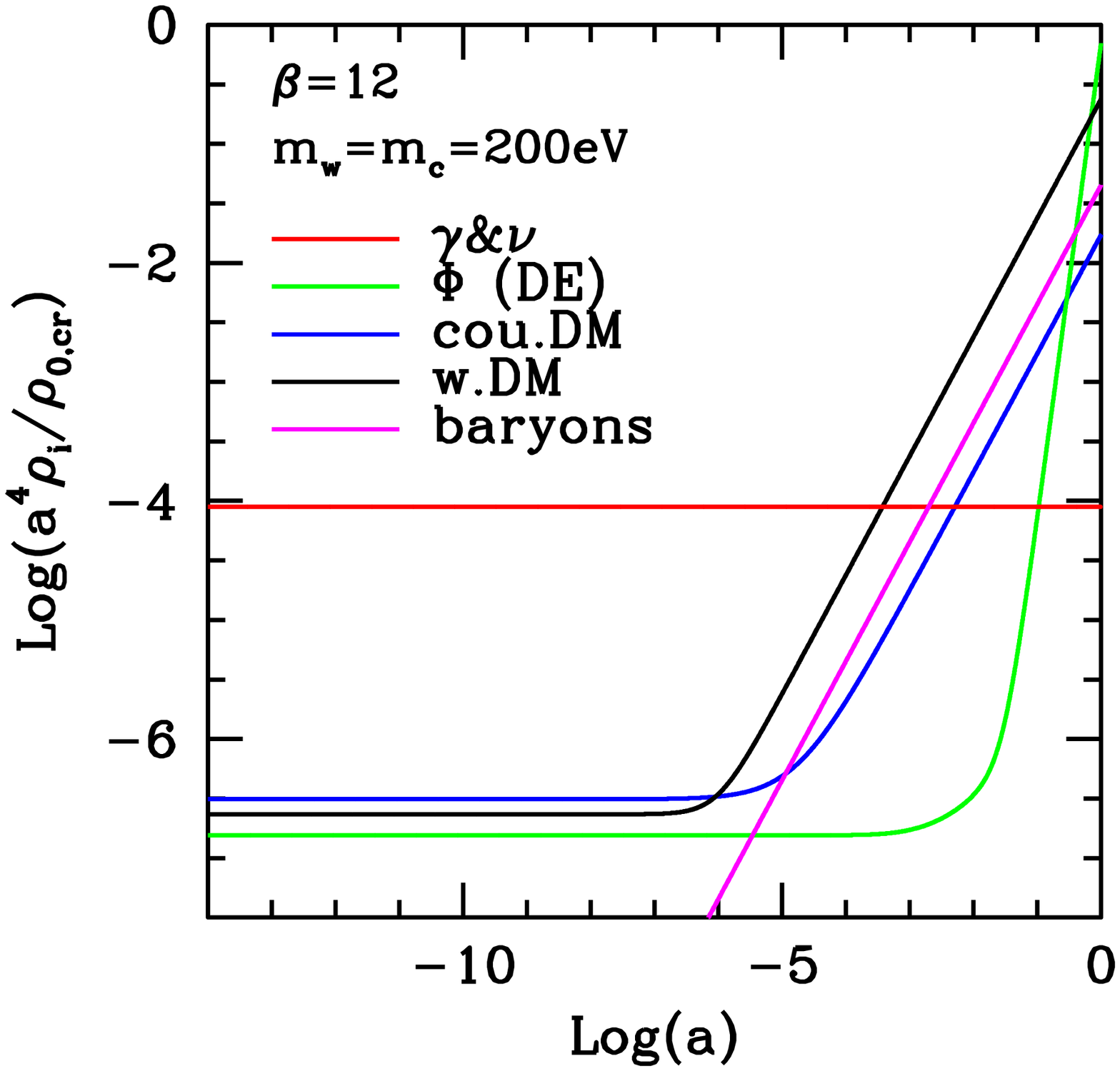}
\includegraphics[height=7.25cm,angle=0]{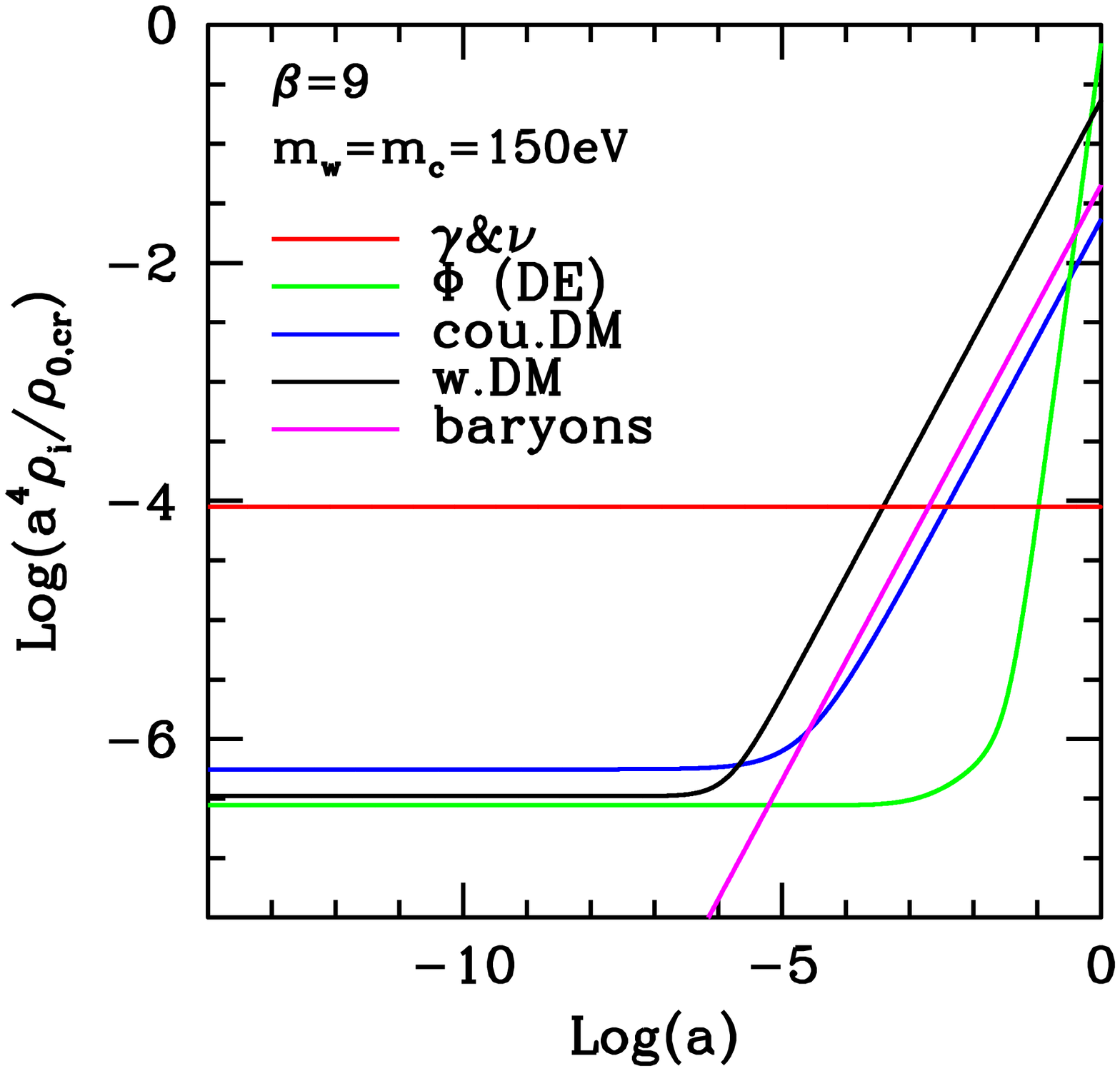}
\includegraphics[height=7.25cm,angle=0]{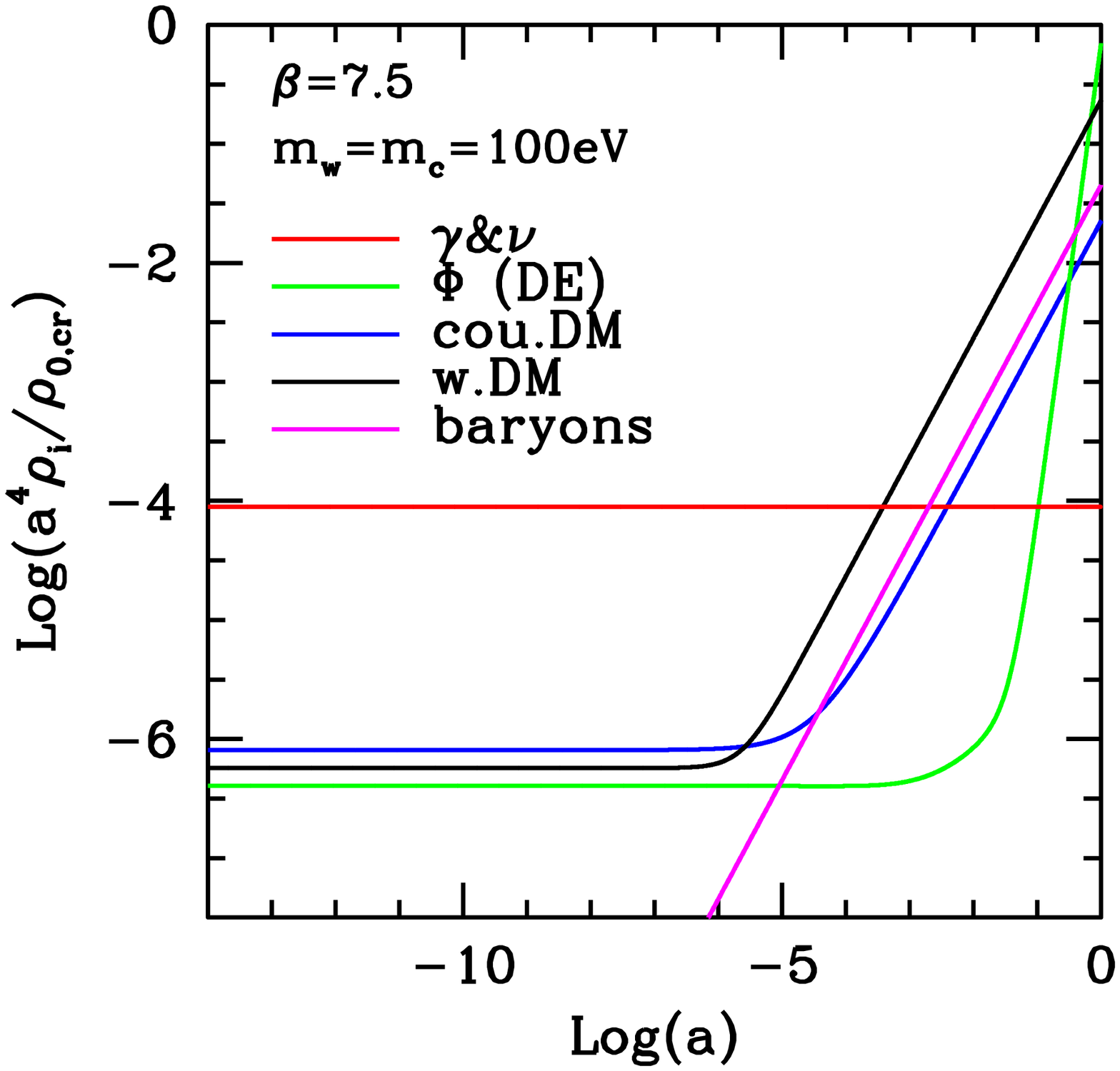}
\includegraphics[height=7.25cm,angle=0]{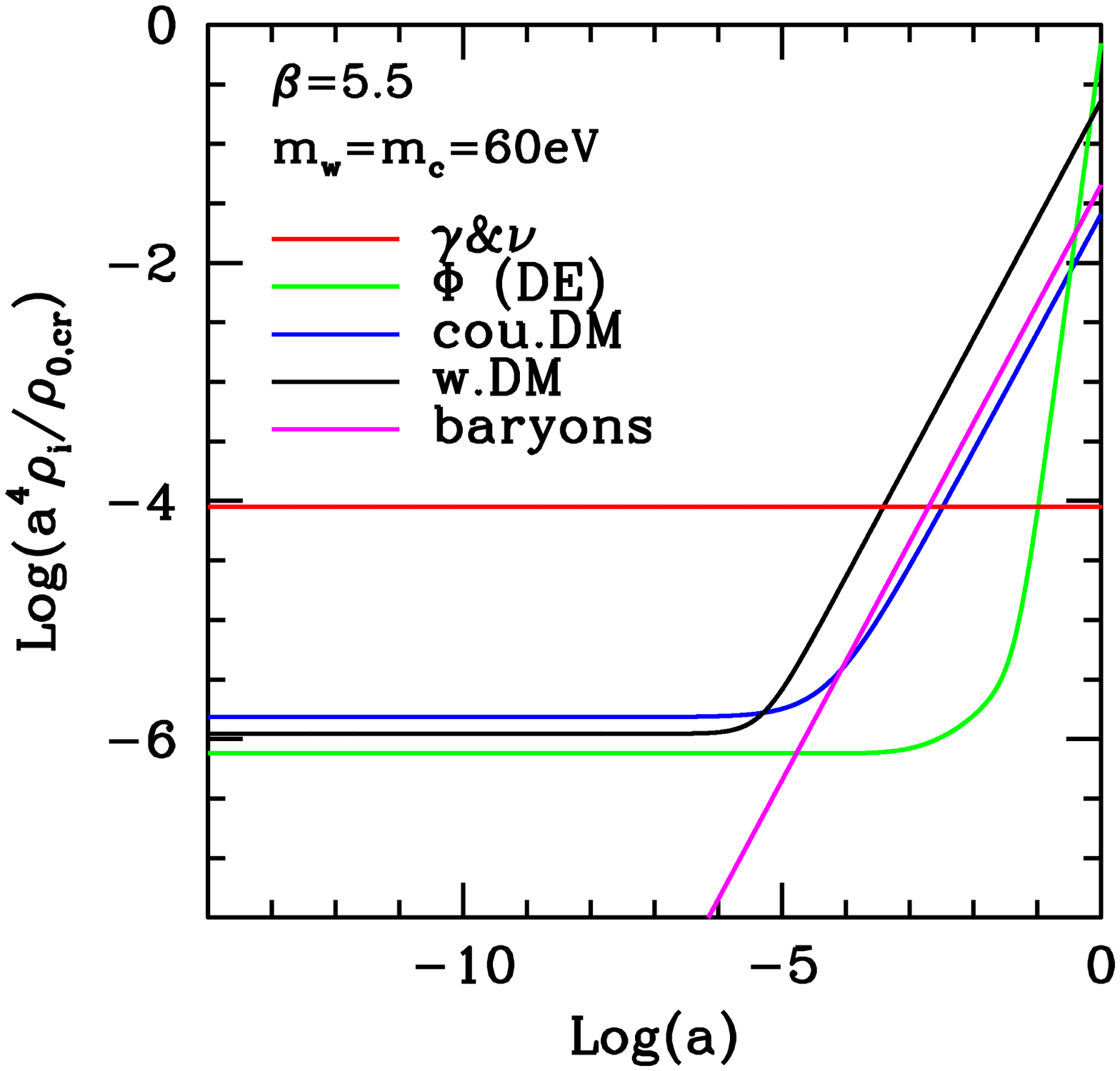}
\end{center}
\vskip -1.truecm
\caption{Density evolution of all components in some SCDEW models.
  Model parameters are shown in the boxes. }
\label{m100-2}
\vskip +.1truecm
\end{figure}

A major decrease of $g$ unavoidably occurs at the quark--hadron (QH)
transition, at a redshift $z_c$, when the cosmic temperature $T_c \sim
170\, $MeV \cite{petre}, as strongly interacting components cease to
be a part of the thermal soup (only pions have a mass $< T_c$,
although slightly so; hadron components however have a proper volume,
and statistical expressions for pointlike particles can be applied to
them only when {\it co}volume does not matter). For a detailed
analysis of the density jump, however yielding a $g$ shift from 61.75
down to $\sim 14.25$, see \cite{universe}. In Figure \ref{rr6} we show
the oscillations of the ratio $(\rho_{c}+\rho_d)/\rho_r$, before the
recovery of the attractor pattern, with radiation density still
including neutrinos and electrons; in the plot electron--positron
annihilation into photons is omitted, so to outline the only effect of
QH transition. A minor quantitative error, made in \cite{universe}, is
also corrected.  The detailed shape of the $(\rho_{c}+\rho_d)/\rho_r$
oscillations slightly depend on specific assumption on the transition
and in some case they can propagate down to primeval nucleosynthesis
(BBN) \cite{universe}.  All lattice QCD results
consistently suggest
a crossover transition.  To obtain the results plotted, however, we
assumed a transition slightly delayed, until the reactions allowing 3
quarks to coalesce within single baryons could perform their duty.
For other assumptions concerning the transition, see again
\cite{universe}. { Incidentally, let us recall that coupled
  components, yielding a fractional contribution to cosmic density
  $\sim 3/4\beta^2$, would have an impact on BBN similar to half
  neutrino species for $\beta \simeq 3.2$; for the greater $\beta$
  values we consider, therefore, there is no appreciable direct
  impact. On the contrary, the residual deviations from a purely
  radiative $a(\tau)$ law, as shown at the r.h.s. of Figure \ref{rr6},
  are approximately half of those due to electron--positron
  annihilation; the impact of such extra deviation is probably
  negligible but was never directly tested}.

Although delayed, the $\rho_c$ growth however means that a common
origin of DM components requires that, at the eve of derelativization,
it must be $\rho_w < \rho_c$.  If $z_d>z_c$, as expected, taking into
account also muon later annihilation, we have an upward jump for
$\rho_c$ by a factor $\simeq 1.63\, .$ Taking also into account other
earlier particle annihilations successive to $z_d$, common origin
requires $\rho_w$ values to lay significantly below $\rho_c$, in the
late CI expansion.

In Figures \ref{m100-2} the evolution of densities in four models
compatible with such conditions are shown.  These Figures extend to
high $z$ without taking into account $g$ variations, as those outlined
in the previous Figure.  Let us then also outline the progressive
increase of $\beta$ with $m_{c,w}$, in the models shown, assuring that
the ratio $\sim 1.63$ (between high--$z$ coupled DM and warm DM
densities) is not badly bypassed (we however neglected the
possibility that the number of spin states of the two components are
different). SCDEW models badly violating the ``1.63'' limit, however,
need a different explanation for the apparent vicinity of high--$z$
values of $\rho_w$ and $\rho_c$. The point, however, is that
low--$\beta$ { and high mass} values seem disfavored by data fits; the
idea of a common nature of the DM components, although fascinating,
seem therefore { uneasy.}

\section{Linear fluctuation evolution}
Fluctuation evolution in SCDEW models was first discussed in
\cite{BM}.  Further points were then outlined in \cite{BMM}, where
models were modified to allow N--body simulations \cite{MMPB}. This
was done by requiring DM--$\Phi$ interactions to rapidly vanish at a
suitable redshift $z< 10^4$. Here, rather than suggesting such {\it ad--hoc}
prescription, we show that  $\beta$ is rapidly cut off to quite low values, 
when the coupled DM field acquires a Higgs' mass. This is one of the main
points of this work.

Initial conditions outside the horizon were treated by \cite{BM} and,
in general, are to be quantitatively modified just for mass scales
\begin{equation}
M = {4 \pi \over 3} \rho_m \left( 2\pi \over k \right)^3
\label{massscale}
\end{equation}
($\rho_m:$ sum of material component densities at $z=0$) close to
today's horizon. In a synchronous gauge, the cosmic metric reads
\begin{equation}
ds^2 = a^2(\tau) [d\tau^2 - (\delta_{ij}+h_{ij}) dx_i dx_j)]~,
\end{equation}
$\tau$ being the universal time, while gravity perturbation is
described by the 3--tensor $h_{ij}$, whose trace is $h$. The system of
differential equations is also to be modified to take into account
variable $\beta$'s and the linear code already used by \cite{BM} and
\cite{BMM} was then modified accordingly.
\begin{figure}[t!]
\begin{center}
\vskip -.5truecm
\includegraphics[height=9.5cm,angle=0]{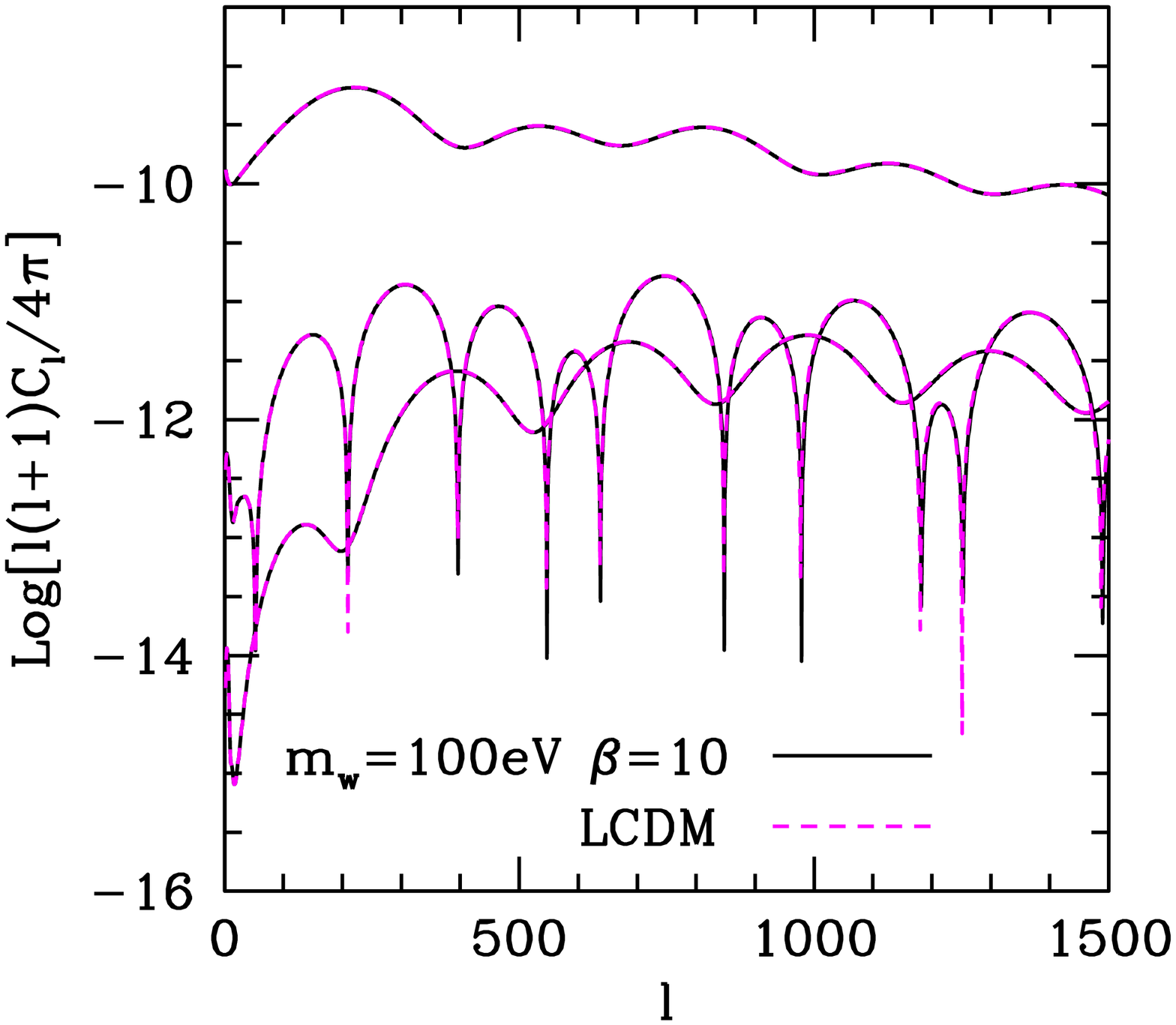}
\vskip -.5truecm
\vskip -.5truecm
\includegraphics[height=9.5cm,angle=0]{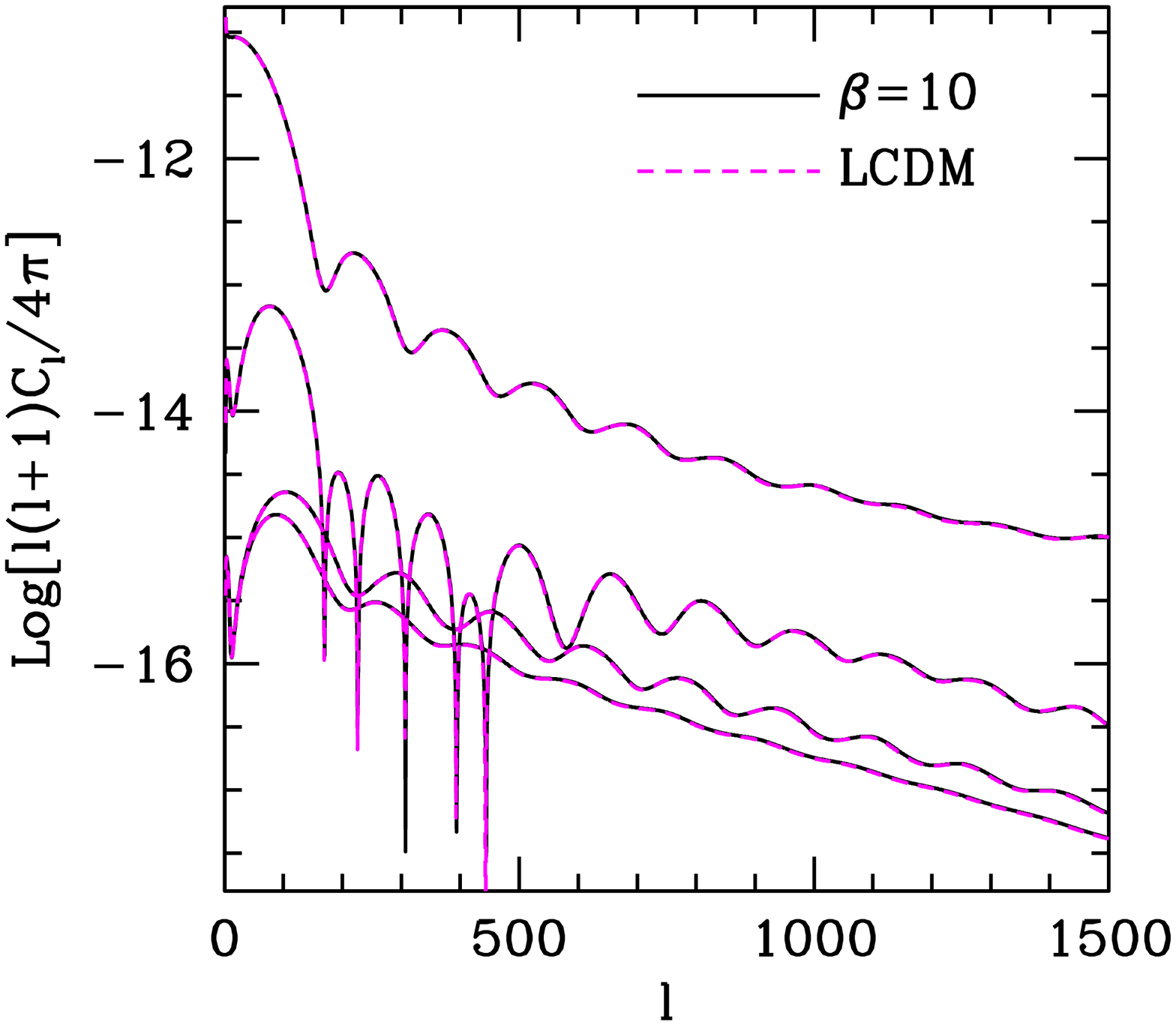}
\end{center}
\vskip -.5truecm
\vskip -.5truecm
\caption{CMB spectra compared; LCDM vs.~SCDEW with $m_{w,c}=100~$eV
  and $\beta=10$. Minimal differencies (see next plot) are essentially
  independent from $m_{w,c}$, and decreasing when greater $\beta$
  values are taken.  }
\label{cl3}
\end{figure}
Let us remind that DE field fluctuations need also to be
considered. 

Let then
\begin{equation}
\phi = \Phi + {b \over m_p} \varphi~,
\label{varphi}
\end{equation}
be the sum of the background field $\Phi$ considered in the previous
Section and a perturbation described by $\varphi$. At the first
perturbative order, the latter $\phi$ component fulfills an equation
containing the term $a^2 V''(\Phi) \varphi$, i.e. including the second
derivative (in respect to $\Phi$) of the self--interaction potential.

This is a critical point we wish to recall, for we prefer to refer to
the $w(a)$ behavior and, therefore, as in \cite{BM}, we replace
\begin{equation}
2V'' = {A \over 1+A} \left\{ {\dot a \over a} {\epsilon \over 1+A}
\left[ \epsilon_6 {\dot a \over a^3} + 2C {\rho_c \over \dot \Phi}
\right] + \left[ {\dot a \over a^3} {\ddot \Phi \over \dot \Phi}
+ {d \over d\tau} \left( \dot a \over a^3 \right) \right]
\epsilon_6 + 2C {\dot \rho_c \over \dot \Phi} \right\}
\end{equation}
with $A$ and $\epsilon$ defined as in eq.~(\ref{kp}). Here $\epsilon_6
= \epsilon-6$, while $\rho_c$ is the coupled DM density.  

Here again the background field $\Phi$ appears just through its
derivatives. The $\varphi$ and $\delta_c$ equations are also to be
added the terms arising from the dependence of $\beta$ (now
$\beta_{eff}$) on $\Phi$ (eq.~\ref{newC}) and the public algorithm
{\sc cmbfast} \cite{cmbfast} is modified accordingly. For this work we
used the version including tensor perturbations.
\begin{figure}[t!]
\begin{center}
\vskip -.5truecm
\includegraphics[height=8.5cm,angle=0]{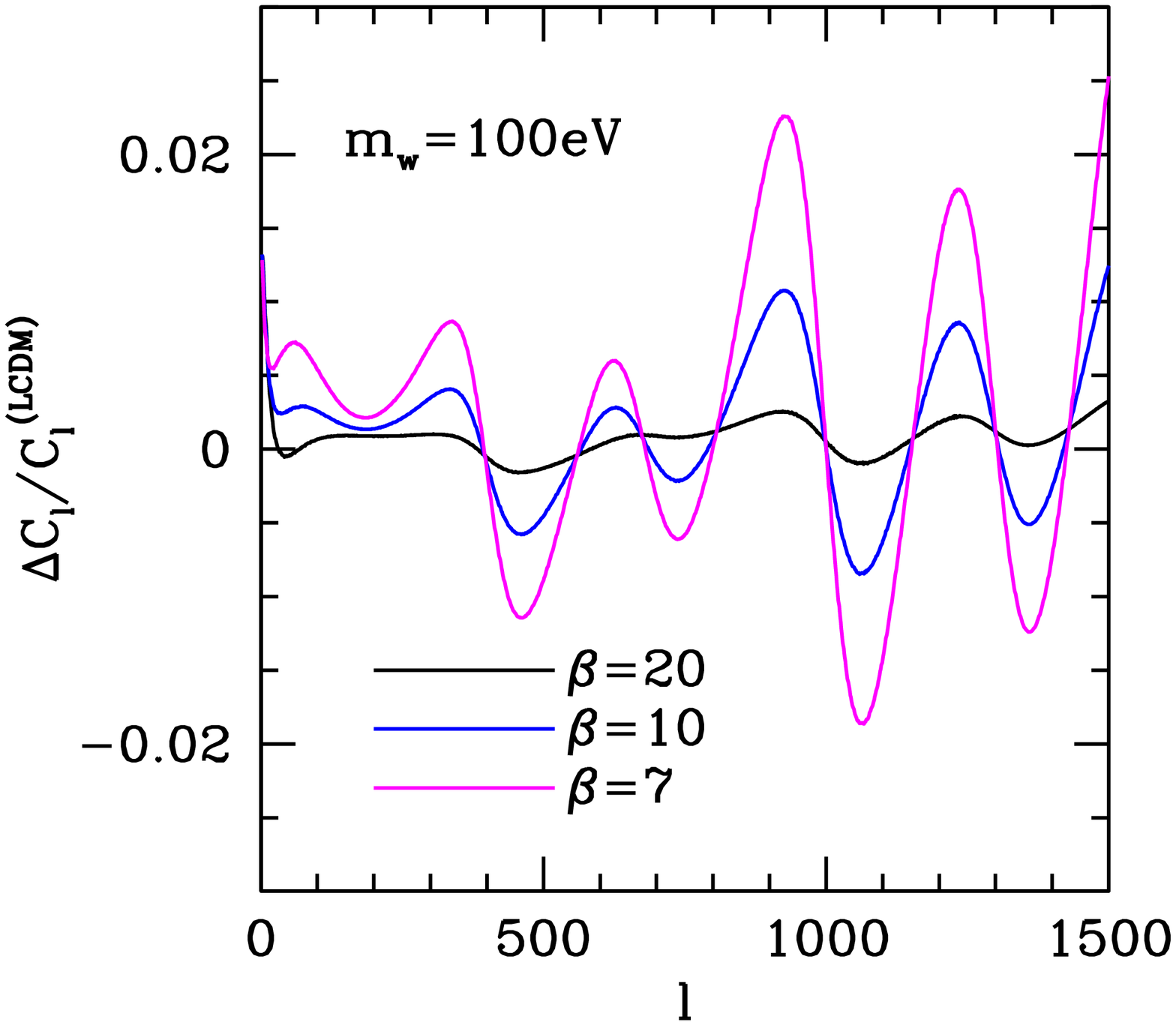}
\end{center}
\vskip -.5truecm
\vskip -.5truecm
\caption{Fractional discrepancies between TT spectra for SCDEW and LCDM. }
\label{Dcl}
\end{figure}
\begin{figure}[t!]
\begin{center}
\includegraphics[height=12.5cm,angle=0]{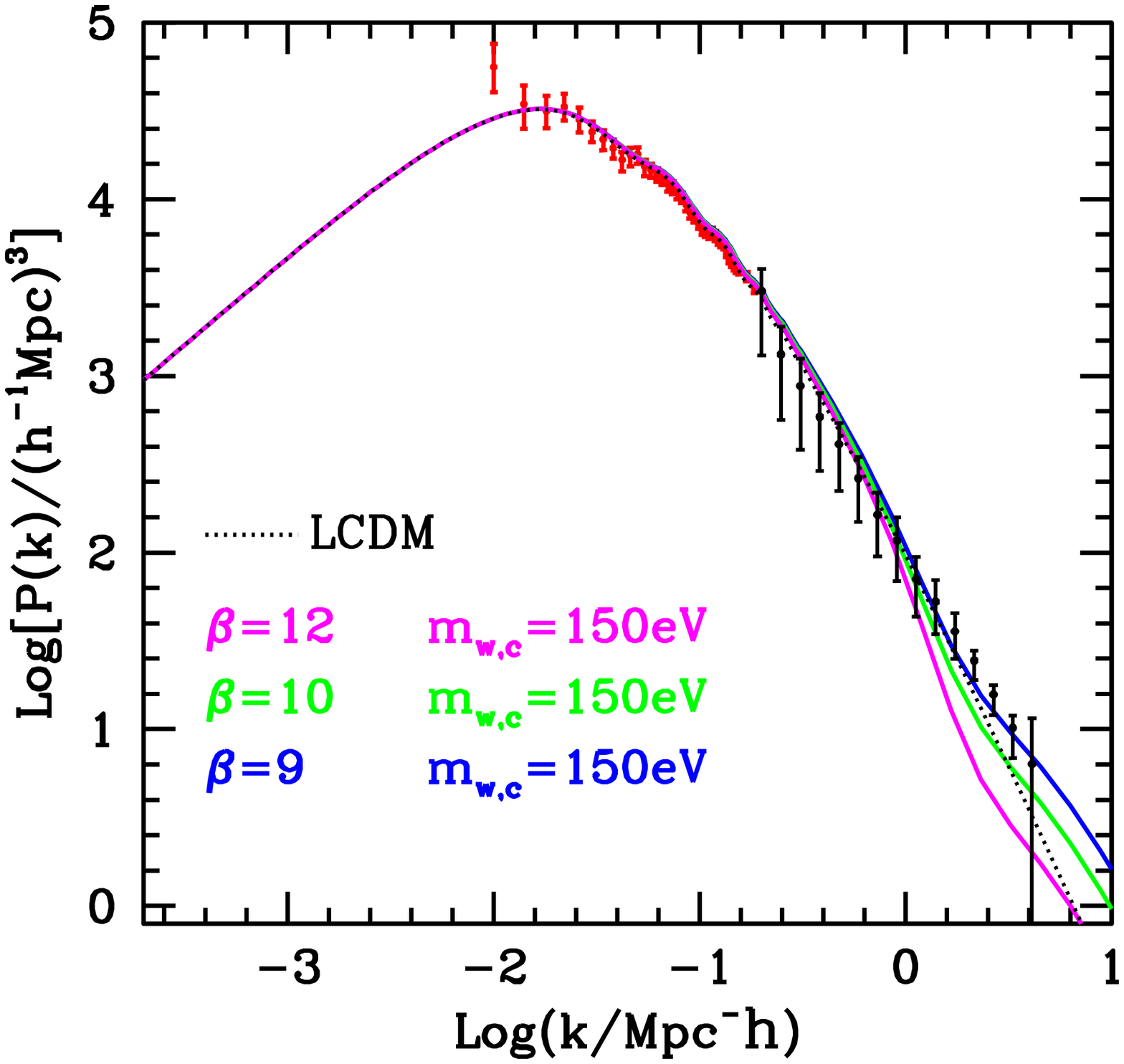}
\end{center}
\vskip -.5truecm
\vskip -.5truecm
\caption{SCDEW and LCDM spectra vs. 2dF (in red) and Lyman--$\alpha$
  (in black) data.}
\label{Qk} 
\end{figure}

\subsection{CMB}
By using modified {\sc cmbfast} we obtain CMB spectra arising from
scalar and tensor perturbations and compare them with LCDM. As it is
known, this model parameters are fitted, first of all, to CMB
data. All through this paper we kept the values of the basic
parameters $\Omega_b$, $\Omega_\Phi$, $h$, $n_s$, $\sigma_8$, etc., to
the same values used for LCDM which, in turn, were the best fit values
obtained in the Planck experiment \cite{planck}. In upper frame of
Figure \ref{cl3} we show the basic TT, TE, EE spectra; in the bottom
frame, the tensor spectra are also shown. This kind of plots are
insufficient to exhibit the tiny differences between SCDEW and LCDM.

In Figure \ref{Dcl} we then show the relative differences between TT
spectra, keeping $m_w=100\, $eV, but considering different couplings.
Differences are stronger for lower coupling, reaching $\sim 2\, \%$
for $\beta=7$ and $\sim 4.5\, \%$ for $\beta=4$ (not shown). Of
course, this does not mean that low--coupling SCDEW cosmologies do not
fit CMB data. Rather, direct tests of model outputs vs available data
should be performed. We however expect some minor variation of the
best fit cosmological parameter values, namely of the overall
fluctuation normalization, well within 2--$\sigma$'s, and no
significant variation of the best fit likelihood.

A possible explanation of the ``better performance'' of large coupling
is the smaller 
contribution of coDM to the total density at a redshift $\sim 1100$,
when most CMB photons undergo their last scattering

\subsection{Linear spectra}
{ Let us then tentatively compare SCDEW and LCDM fluctuation
  spectra at $z=0.$ We shall take into account also data from the 2dF
  redshift survey and Lyman--$\alpha$ analysis \cite{cole, zaroubi},
  so to compare SCDEW--LCDM discrepancies with observational trends
  and errors.  Seeking} the best fit values of the SCDEW parameters
(e.g., $m_{w,c}$ and $\beta$) by using { available} datasets, is
out of our scopes in this analysis.
\begin{figure}[t!]
\begin{center}
\includegraphics[height=7.5cm,angle=0]{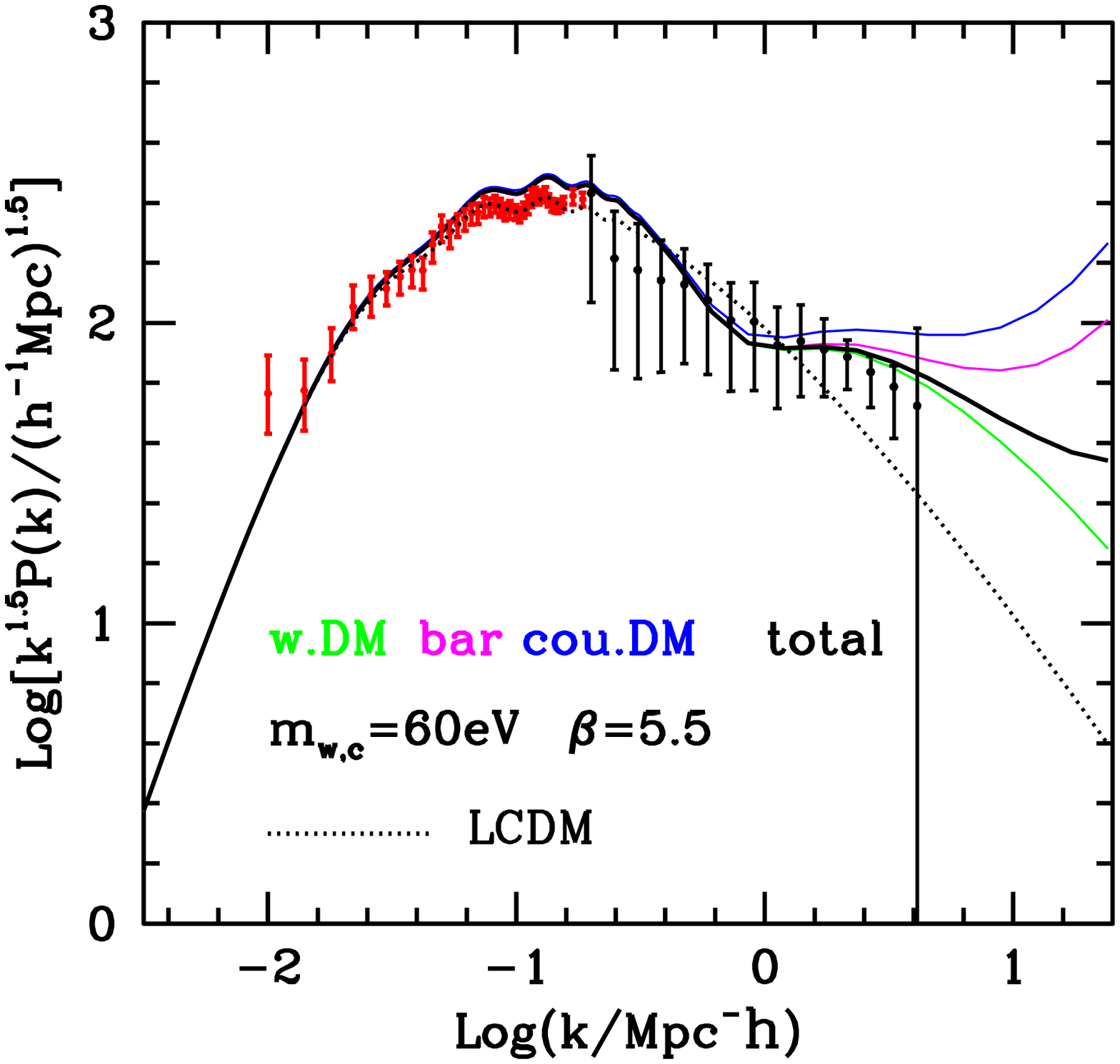}
\includegraphics[height=7.5cm,angle=0]{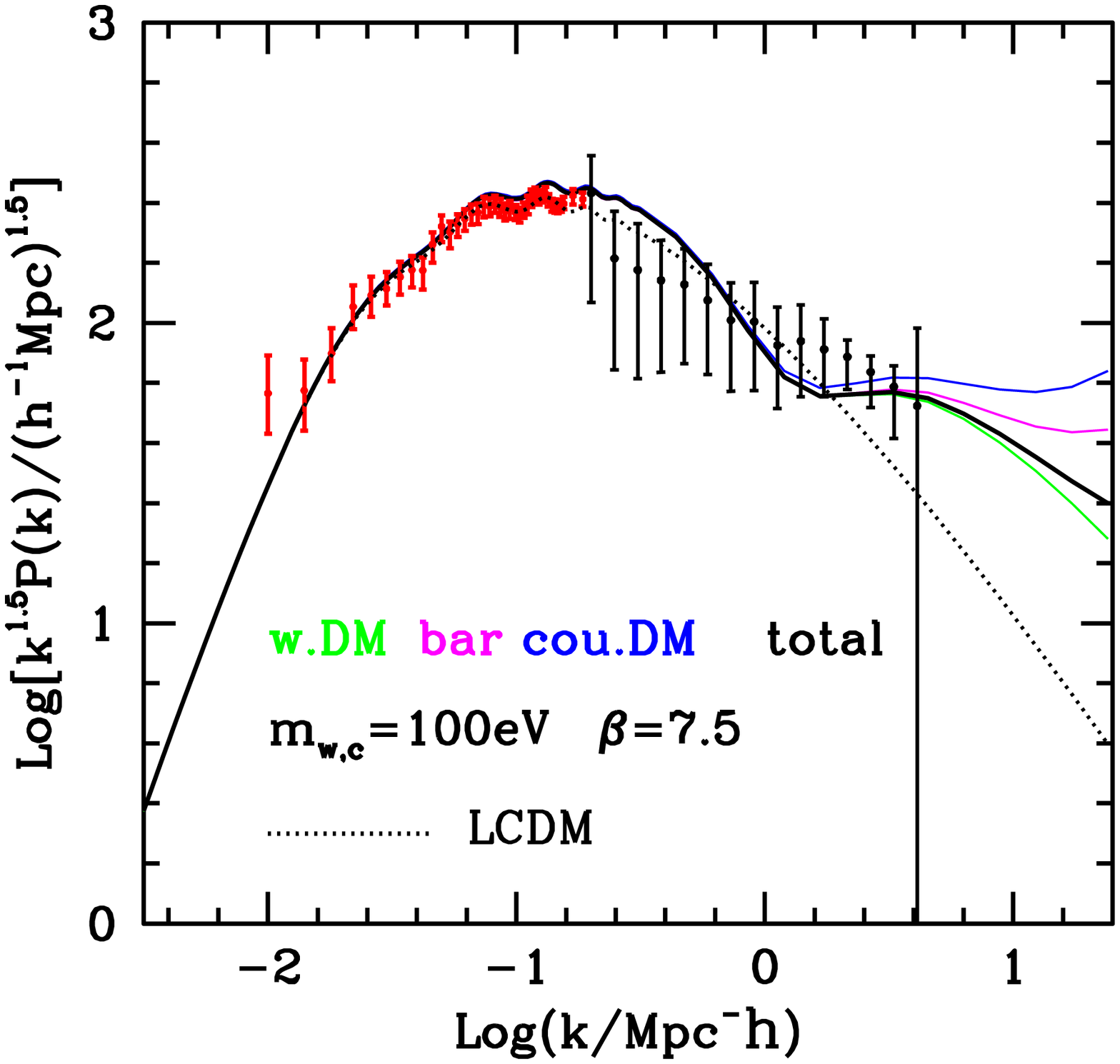}
\includegraphics[height=7.5cm,angle=0]{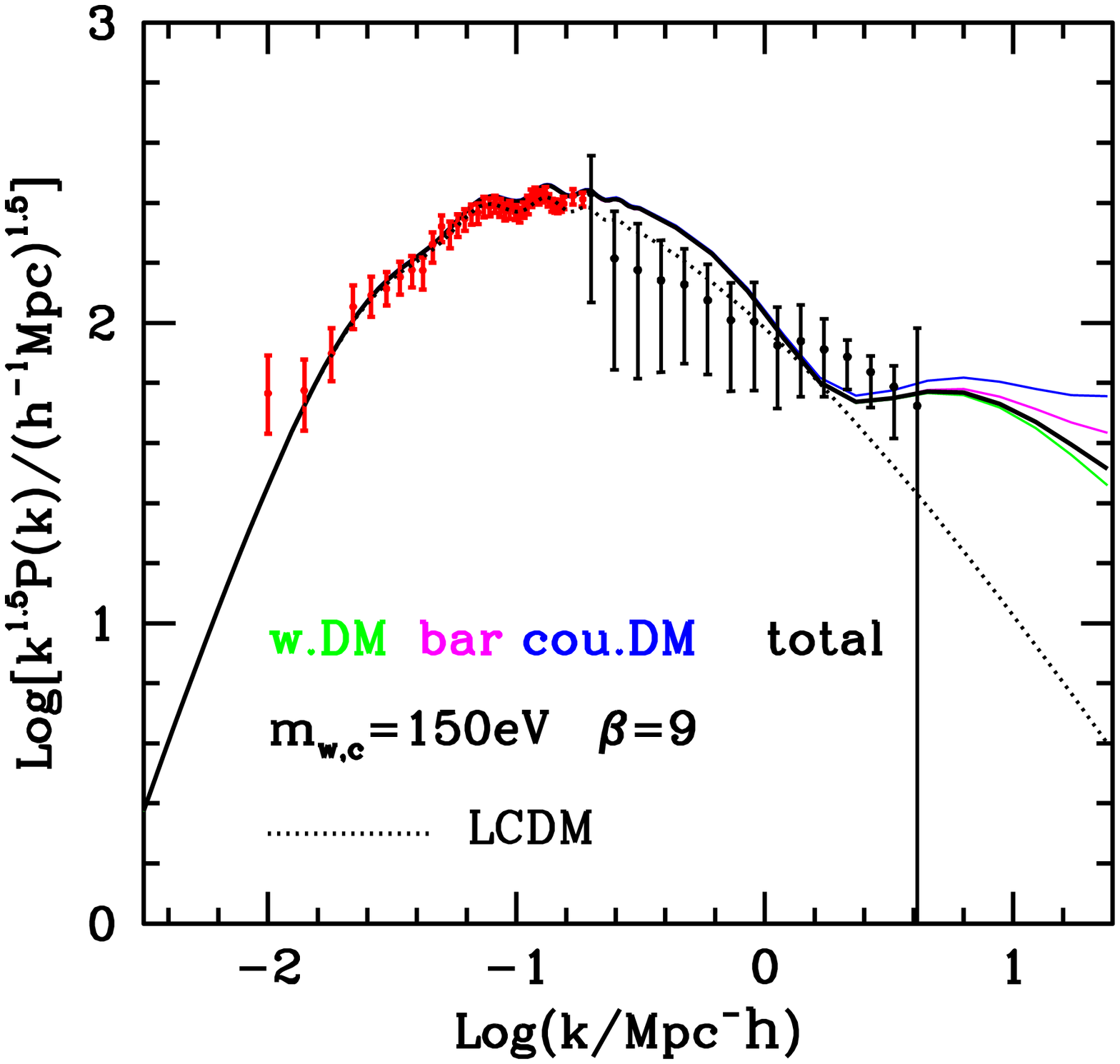}
\includegraphics[height=7.5cm,angle=0]{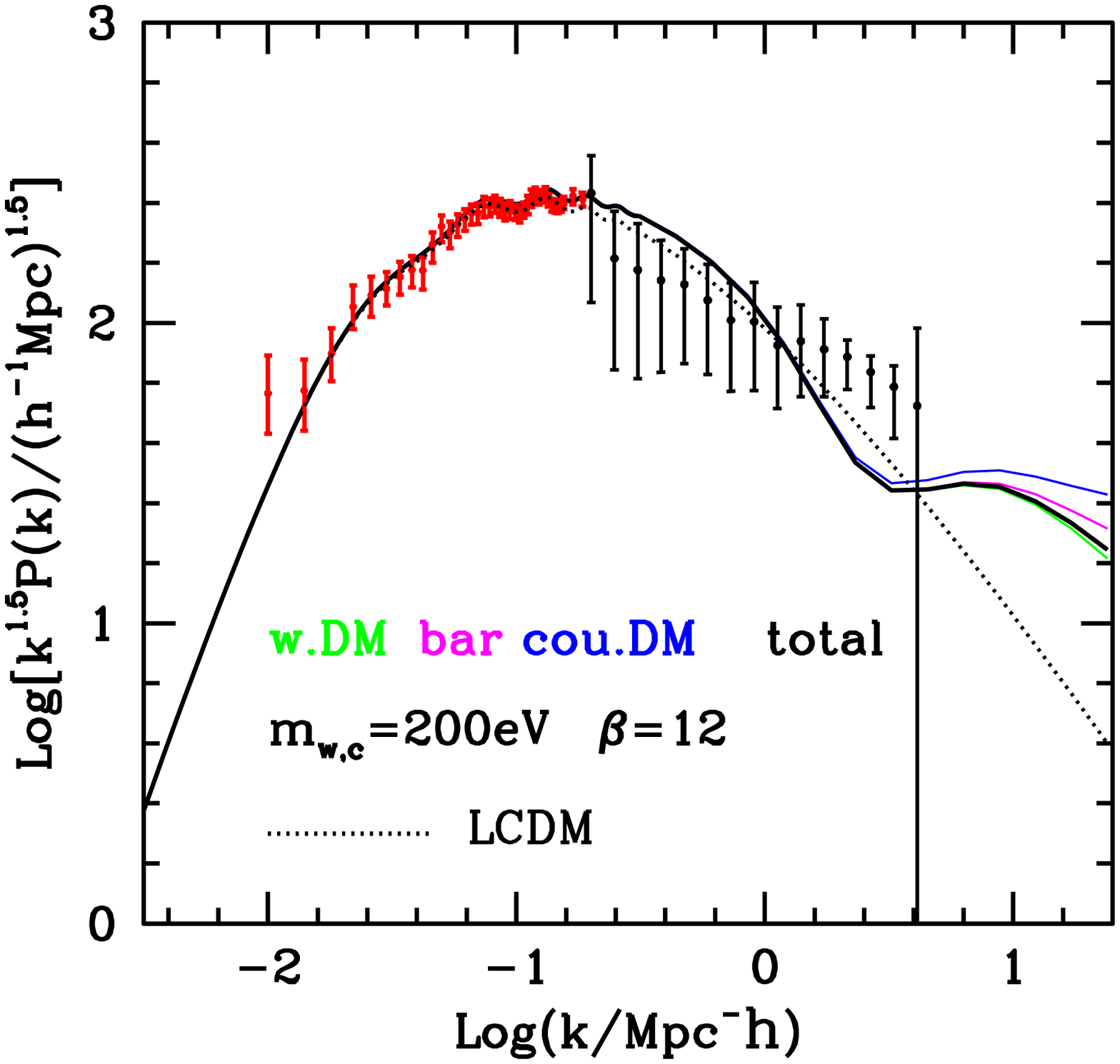}
\end{center}
\vskip -.5truecm
\vskip -.5truecm
\caption{SCDEW and LCDM spectra vs. 2dF (in red) and Lyman--$\alpha$
  (in black) data. Spectra are obtained from linear programs up to
  fairly large $k$ values. }
\label{QkLR}
\end{figure}
{ Let us also outline that the physical mechanisms yielding SCDEW
  spectral shapes are discussed in the next Subsection.}

In Figure \ref{Qk} we then show the spectra for three SCDEW models and
LCDM vs.~data. The aim of this Figure is to illustrate how models
depend on the value of the $\beta$ parameter. Henceforth we keep
$m_{w,c} \equiv 150\, $eV while the couplings $\beta = 9$, 10, 12 are
considered. At large scale ($k \lesssim 0.05\, h\, $Mpc$^{-1}$)
SCDEW--LCDM discrepancies are small and data are similarly fit by LCDM
and SCDEW. For greater $k$ values, however in the region where
Lyman--$\alpha$ data are available, the fit appears better for smaller
$\beta$'s. At still greater $k$ values, SCDEW spectra tend to exceed
LCDM and to do more so for smaller $\beta$ values.

The excess power for $\beta \sim 9$ concern scales
which, in the present epoch, have reached the non--linear
regime. Accordingly, it could indicate an early formations of
galaxies, groups and small clusters.
Let us however outline that the fit of $z=0$ spectra with
Lyman--$\alpha$ data, concerning large--$z$ systems, { is slightly
  biased by the use of fluctuation evolution in a LCDM cosmology}.

{ Further comparisons are provided in Figure \ref{QkLR}, by
plotting $k^{3/2}P(k)$, instead of $P(k)\, $ for the same SCDEW
cosmologies of Figure 4}
(incidentally, a comparison between Figure \ref{Qk} and
\ref{QkLR} shows the extraordinary impact of the way how spectra are
shown, in appreciating spectral distorsions).

The spectra shown,
up to $m_{w,c}=150\, $eV, {are quite close to LCDM and} meet
observational errorbars, {at least, as well as it; but also the
  last model keeps inside 3 $\sigma$'s, while we should also}
recall 
how Lyman--$\alpha$ data are extrapolated { to $ z=0\, .$}

These plots { also}
allow us to appreciate that different components still exhibit
different transfer functions at $z=0$, although at large $k$ values,
an effect stronger for smaller couplings. Another point is that, in
order that models approach data, masses and $\beta$ must be
simultanously increased. A greater mass value causes an earlier
weekening of the $\Phi$--$\psi$ coupling, so that the final amplitude
of coupled DM spectra is decreased. 

It is however clear that the most significant discrepancies between
SCDEW and LCDM shall mostly concern non--linear scales, so that
discriminating between the two cosmologies requires a difficult
interplay between spectral differences and assumptions about baryon
physics.


\subsection{Linear and non--linear fluctuation evolution}
\begin{figure}[t!]
\begin{center}
\vskip -.5truecm
\includegraphics[height=7.4cm,angle=0]{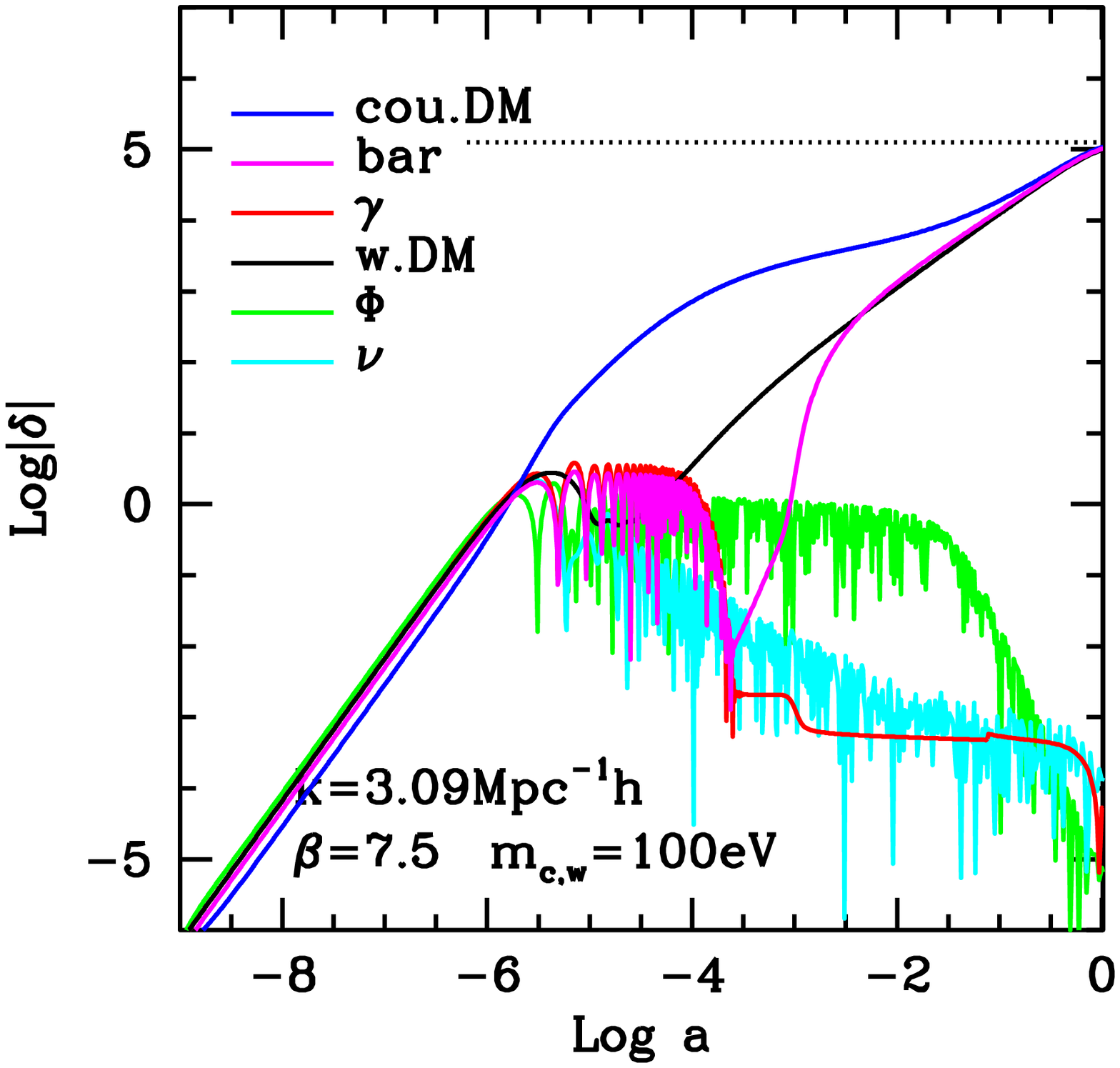}
\includegraphics[height=7.4cm,angle=0]{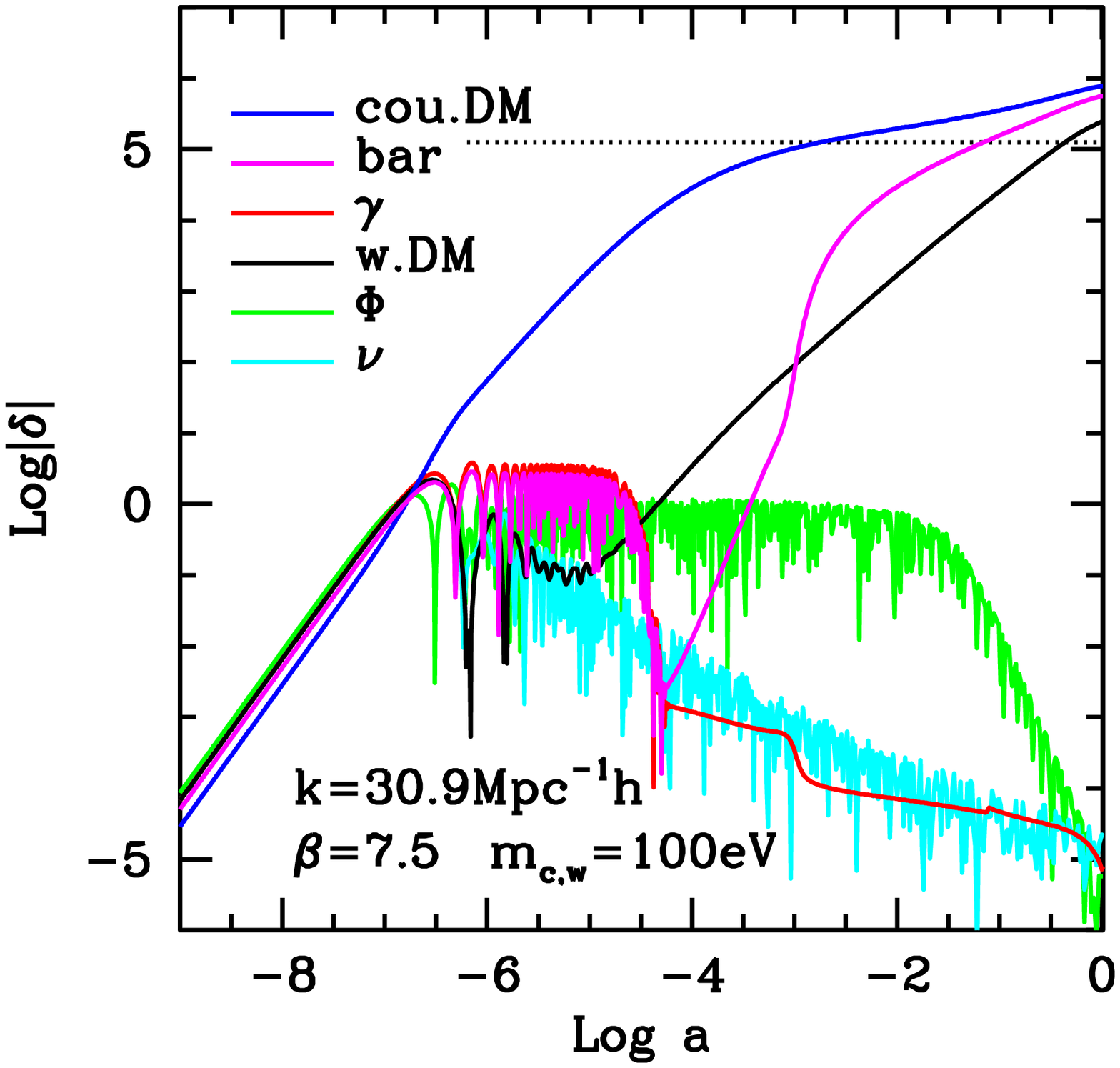}
\includegraphics[height=7.4cm,angle=0]{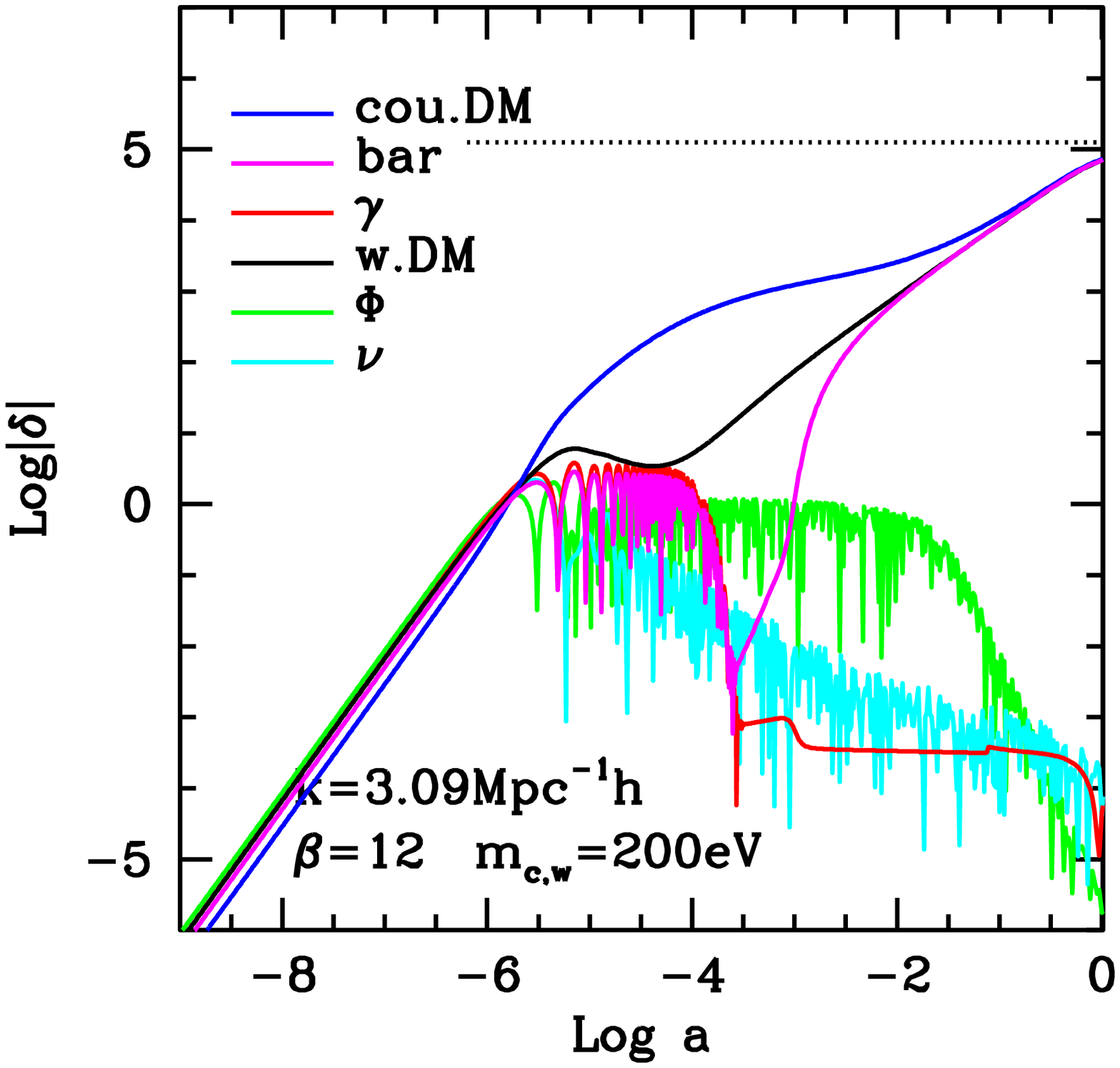}
\includegraphics[height=7.4cm,angle=0]{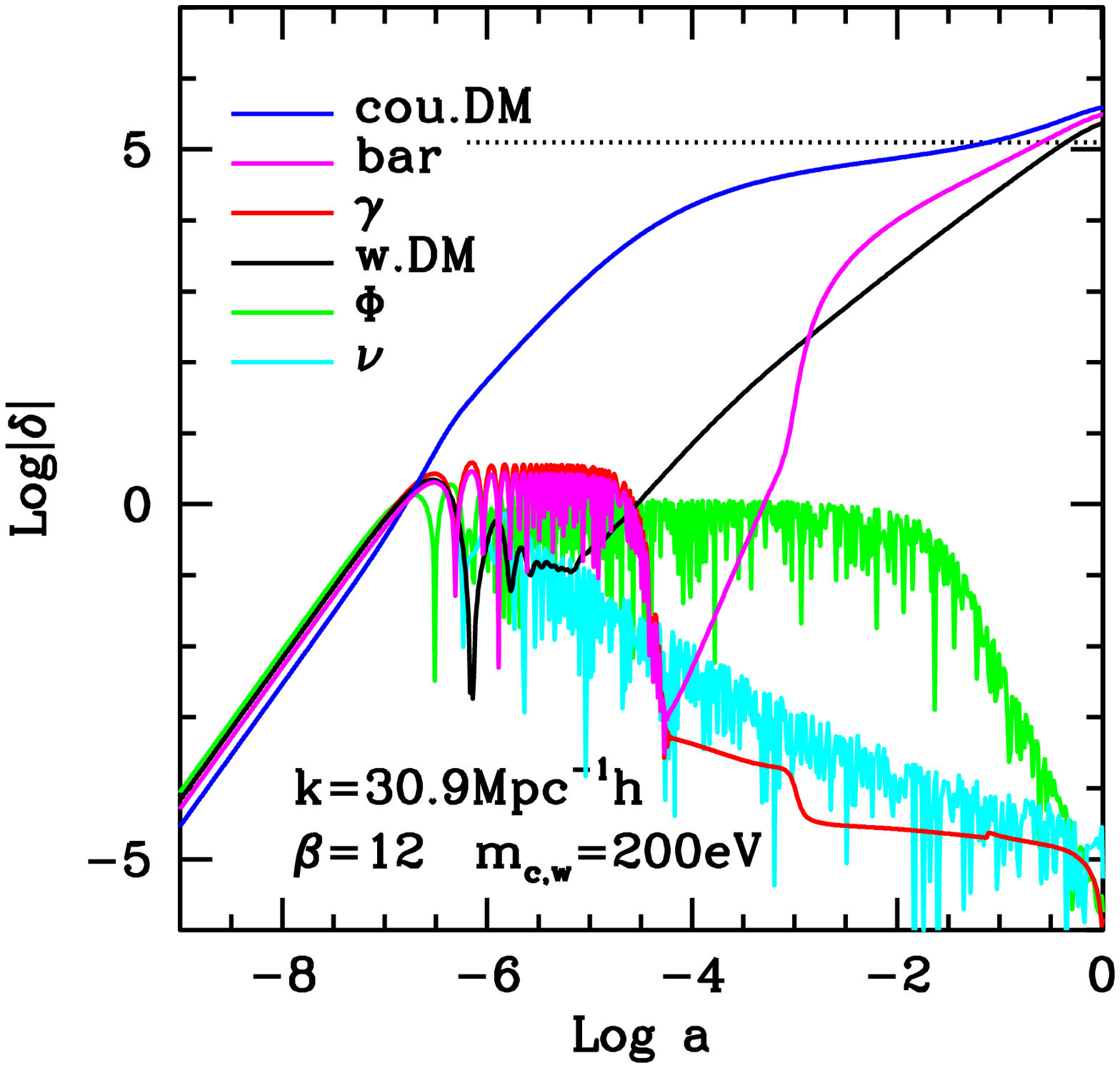}
\end{center}
\vskip -.5truecm
\vskip -.5truecm
\caption{Linear evolution of fluctuations in SCDEW models.  Magenta
  and green curves show $\gamma$ and $\nu$ evolution,
  respectively. Matter components are colored as shown in the
  frames. Gravitational and $\Phi$ fields are omitted. The $k$ scales
  considered are precisely 30.8935 and 3.08935$\, h$Mpc$^{-1}$,
  corresponding to 0.8 and 8$\, h^{-1}$Mpc or to $\sim 1.79 \times
  10^{11}$ and $1.79 \times 10^{14} h^{-2}M_\odot$, respectively.}
\label{fluevo}
\end{figure}

In Figure \ref{fluevo} we show the evolution of fluctuations with
comoving wavelengths $\lambda_0 = 8$ and 0.8$\, h^{-1}$Mpc
(approximately $k=3.09$ and 30.9$\, h$Mpc$^{-1}$) for the above models
with $m_{w,c}=100$ and 200$\, $eV.

{ These plot show the basic mechanism allowing models with
low mass WDM to yield structure down to quite small mass scales.
This is due coDM fluctuations growth
also during the so--called {\it stag--flation} epoch, when uncoupled
DM fluctuations, even if made of fully non --relativistic materials,
are stuck to their amplitude at the horizon entry.

This effect can be easily understood, at least in the
non--relativistic regime. In fact, as outlined in \cite{MQMAB}, in
such regime 
coupling is equivalent to}
an apparent increase of the effective gravitational constant, when
forces acting between coupled DM particles are considered. In such
interactions, the gravity constant $G$ is to be replaced~by
\begin{equation}
G^* = \gamma_{eff}\, G ~~~~{\rm with} ~~~ \gamma_{eff} = 1 + 4
\beta_{eff}^2/3~.
\label{G*}
\end{equation}
Other gravitational forces, even involving coupled DM particles, are
still ruled by the ordinary $G$. Furthermore, while the dynamical law
${\bf f} = {\bf p}'$ continues to hold 
(here {\it prime} indicates differentiation with respect to time $t$),
the particle mass decrease in time yields
\begin{equation}
m {\bf v}' = {\bf f} +  m{\bf v} \left| m' / m \right|~,
\label{dynamics}
\end{equation}
so that, besides of gravitational forces, coupled DM particles receive
an {\it extra push} due to the second term at the r.h.s., sometimes
unappropriately called {\it friction} term.

Let us also outline that the $G$ boosting factor $\gamma =
1+4\beta^2/3$ partially cancels the fact that coupled DM has a density
parameter $\Omega_c \simeq 1/2\beta^2$, 
{ so that the factor}
\begin{equation}
\label{newG}
\gamma \Omega_c = 1/2\beta^2 + 2/3
\end{equation}
multiplying $G = m_p^{-2}$, almost $\beta$--independent for large
couplings, { apparently indicates that coDM fluctuations
  evolve just as fluctuations with an amplitude reduced by a factor
  $\sim 2/3$, but concerning the whole critical density. But even the
  factor 2/3 is widely compensated by the {\it extra push}, so
  yielding the observed fast growth rate.}


The dotted horizontal line, in Figures \ref{fluevo}, shows where
$\delta = 1$ if $\sigma_8 = 0.82\, .$ This confirms that, at the scale
of 8$\, h^{-1}$Mpc, SCDEW models and LCDM are still fully compatible.
Results are different at 0.8$\, h^{-1}$Mpc. Here, in both models,
coupled DM fluctuations attain non--linearity earlier than other
non--relativistic components, while different components did not yet
converge onto a single $\delta$ value,
{ non--linearity is reached at different redshifts in the two
  models:}
in the case $\beta=7.5$, { infact,}
coupled CDM is already non--linear at $z \sim 1000$, for a scale of
$\sim 10^{11} h^{-2} M_\odot$. On the contrary, in the model with
$\beta=12$, { $\delta_c$ attains non--linearity at $z \sim 100\, $.
}

{ This has an evident impact on the techniques used
to analyse non--linear evolution (e.g., N--body simulations) and is due
to SCDEW models relying on coDM fluctuation persistence to revive
warm--DM fluctuations --at $z < z_{der} \sim 6 \times 10^{\, 5}
(m_w/100\, {\rm eV})^{4/3} (\Omega_w h^2)^{-1/3}$-- and baryon
fluctuations, when they decoupled from $\gamma$'s. This is the key
effect, overcoming the former erasing of fluctuations in such
components}
because of free streaming and sonic wave damping, respectively.
{ However, if coDM fluctuations enter non--linearity too early, there
  is a numerical problem to evaluate fluctuation evolution and,
  possibly, a physical problem.}

{ Clearly, coDM non--linearities depend on the mass scale. Over very
  large mass scales, average fluctuations have not yet entered
  non--linearity even today. A safe boundary is $\simeq 2 \times
  10^{14} h^{-2} M_\odot$, as shown by the plots in the first column
  in Figure \ref{fluevo}. At any smaller mass scales the redshift
  where coDM attains non linearity depends on model parameters, being
  smaller for greater~$\tilde \mu$. When this occurs, linear
  predictions on all components are unsafe, although it should be
  reminded that $\delta_c>1$ does not mean that the overall
  fluctuation $\delta > 1$. Furthermore, as shown in the second
  associated paper, the top density contrast of the coupled component
  could be just $\cal O$$(20)$, so hardly causing, by itself, a
  $\delta > 1$.

It is however evident that, over scales $\lesssim 10^{10}$--$10^{11}
h^{-2} M_\odot$, non--linear effects can occur earlier in SCDEW than
in LCDM. As these scales are non--linear today, a greater spectral
amplitude essentially means an earlier system formation. As a matter
of fact, however, the rise of observable structures over such scales
depends on the growth of baryonic fluctuations, whose dynamics is
ruled by hydrodynamics and dissipative effects.

Such earlier non--linearity has also an impact on the formation of
unconventional structures, e.g. large Black Holes; although one could
suggest that these expected features meet recent findings at high--$z$
\cite{pallosobra,pacucci}, predictions are hard to schematize and our
only claim here is that these features cannot be considered as a model
difficulty.  This is however the most intricate scale range to treat
and ends up around a mass scale $M_{min} \simeq 10$--$100\, h^{-2}
M_\odot$, as we show below.

In the second related paper \cite{BM2}, we actually considered a spherical
density enhancement entering the horizon so early to turn around and
virialize before $z_{der}$. After determining the time needed to reach
virialization, we discover a very peculiar feature: that virialized
structures will eventually dissipate. The time elapsing from $\delta_c
\sim 10^{-3}$ (redshift $z_3$) and virialization makes $z_{3} \simeq
100 \times z_{vir}$. Moreover, for average size fluctuations, $z_3
\sim 0.1 \times z_{hor}$. Accordingly, while in LWDM models
fluctuations are able to survive only if their size is reached by the
horizon after $z_{der}$, the presence of coupled--DM in SCDEW models
shifts the critical redshift from $z_{der}$ to $\sim 10^4 z_{der}$, at
least, so lowering by $\sim 12$--14 orders of magnitude the mass scale
of the minimal surviving WDM fluctuation.

At $z_{eq} \sim 2 \times 10^4$ (matter--radiation equality), the
horizon mass is $\sim 10^{17} h^{-2} M_\odot$; then, at $z_{der} \sim
30$--$50 \, z_{eq}$, the horizon mass has lowered to $\sim 10^{13}
h^{-2} M_\odot$; a further jump along $z$ by a factor $10^4$, finally
lowers $M_{min}$ to $\sim 10\, h^{-2} M_\odot$. This
semi--quantitative estimate however assumes an almost instant
dissolution of fluctuations after virialization. If the time taken by
such process is long, $M_{min}$ could be even smaller.  On the
contrary, the time scales deduced through the study of spherical
fluctuations, are just approximate. The above conservative estimate
$M_{min} \simeq 10$--$100\, h^{-2} M_\odot$ is meant to cover us
against any such risk.

Altogether, while SCDEW models substantially overlap LCDM predictions
over a wide range of scales, their comparison with LCDM becomes harder
over scales $\lesssim 10^9$--$10^{10} h^{-2} M_\odot$.  { There is
  however a wide range of SCDEW models which can be expected to
  approach cosmological observables over the full range of
  scales. Discriminating between LCDM and these models requires
  further work involving non--linear baryon physics.}

\section{Discussion}
In previous works, we considered Strongly Coupled DE models either
with coupling persisting down to $z=0$, or by introducing an {\it
  ad--hoc} fading of coupling at a fixed redshift. The latter option
allowed us to perform N--body simulations without needing suitable
modifications of standard programs. Here below, we shall provide some
further comments to these simulation results.

In this work we { however} adopted a third option, already
envisaged in the Appendix of \cite{BMM}, which appears simple and
highly effective. Most of the detailed results of this paper cannot
prescind from its use.

In order to understand its features, let us start from recalling that
the distance between the Planck scale $m_p$ and the Higgs scale $m_H$
is a phenomenological fact lacking explanation. One of the elements in
favor of Super--Symmetries, e.g., is that they allow to preserve such
distance against radiative corrections; but the very distance is an
{\it ad--hoc} prescription of the standard model. The masses acquired
at the $m_H$ scale are also spread over various orders of magnitude,
ranging from heavy quark and intermediate boson masses, themselves
$\cal O$$(m_H)$, down to the unknown neutrino masses $\ll 1\, $eV.
Attempts to go beyond the standard model for fundamental interactions
are also motivated by the hope to get rid of such a mess of
parameters.

Cosmology adds its own contribution to this anthology. In LCDM models,
the exit from radiation dominated era is set by nonrelativistic
particle densities; both baryonic and DM densities being set by the
product of tuned particle masses and number densities. Therefore,
assuming a twofold DM nature, at first sight, furtherly worsens an
already intricated situation. In spite of that, in the literature,
several authors did consider such option on purely phenomenological
bases.

On the contrary, the world picture described by SCDEW models seems to
ease the problem. Even letting apart the option of a common origin of
coupled and uncoupled DM components, both of them require just similar
number densities close to $\gamma$'s or $\nu$'s. Once a single DM
Higgs' mass $m_w = \cal O$$(100\, $eV) is assumed for coupled and warm
DM quanta, the total variable mass of the former particles reads
\begin{equation}
\label{meff1x}
m_{eff} = \mu \exp[-(b/m_p)(\Phi-\Phi_p)] + m_w
\end{equation}
while, more significantly, an exponential cut--off of the coupling
constant follows. The point is that, once the coupled DM component
acquires a Higgs' mass, also its coupling fades, according to the law:
\begin{equation}
C_{eff} = {C \over 1 + (\mu/m_w) \exp[C(\Phi-\Phi_p)]}~.
\label{beteffx}
\end{equation}
Let us recall that the behavior of $\Phi-\Phi_p$ is fixed by dynamical
equations, so that the choice of $C $ and $m_w$ also defines the
residual today's (very weak) coupling.

Accordingly, the presence of coupling and a twofold DM component, in
SCDEW, rather than complicating the cosmological scenario, already
appears as a sort of rationalization. This, however, is not the only
point in support of SCDEW cosmologies. The persistence of the DE field
$\Phi$ with an energy density constantly $\cal O$$(10^{-2}\rho_{cr})$
through various redshift decades, suggests that it played a role at
both ends of this stationarity era which, in SCDEW cosmologies,
replaces the ordinary radiative era; henceforth, we can expect that
such conformally invariant expansion is triggered by inflation end and
reaches iself an end as a consequence of Higgs mass acquisitions.
Meanwhile, the $\Phi$ increase is just logarithmic, so that its
present value is $\cal O$$(60\,  \Phi_p)$. Even more significantly, in
the potential $V(\Phi)$, that we refrained from detailing (and is
responsible for the transition of $\Phi$ from its early kinetic
behavior to the contemporary potential behavior), there should appear
suitable constants which also grew logarithmically, according to the
renormalization group equations, from the inflationary to the present
era. Accordingly, one could attempt to build specific models where the
same potential $V(\Phi) $ exceeds the kinetic energy $\dot
\Phi^2/2a^2$, when $\Phi$ is large enough, either during inflation or
at the eve of the present epoch.

The attractor nature of the conformally invariant expansion was
illustrated here by following the recovery of the tracking regime when
the number of spin degrees of freedom of the primeval {\it thermal
  soup} drastically changes. The example taken was the transition of
strongly interacting matter from the early quark--gluon plasma to a
hadron gas. In this connection we also evaluated the expected
relevance of some (mild) consequences on primeval BBN.

Among the points illustrated in this work, there are CMB spectra due
to both scalar and tensor perturbations. Here we found just small
discrepancies from LCDM, however greater for lower values of the
coupling constant $\beta$ (for some multipoles up to $\sim 2\, \%$ if
$\beta \sim 7$ and $m_w = 100\, $eV).

At variance from previous expectations, in this work we argue that the
coupled DM component of very small scale fluctuations, entering the
horizon early enough, undergo an authonomous evolution, ending up with
their total dissipation. The whole process is intrinsically different
from the well know {\it free--streaming} of relativistic particles
from early inhomogeneities, as soon that their size is reached by the
horizon scale. The process we envisage proceeds through two basic
stages, described in more detail in a second associated paper \cite{BM2}: 
(i) The
effective self--gravity of coDM, boosted by its very coupling, causes
coDM fluctuation to grow, even when the other components form sonic
waves or freely stream. Such growth reaches a non--linear regime, so
yielding a density contrast between coDM fluctuations and average coDM
density $\sim 25$--26, at virialization. (ii) But coDM particle mass
still decreases and virialized configurations are unstable: particle
orbits necessarily widen, finally destroying the initial inhomogeneity.
\begin{figure}[t!]
\begin{center}
\vskip -.1truecm
\includegraphics[height=10.8cm,angle=0]{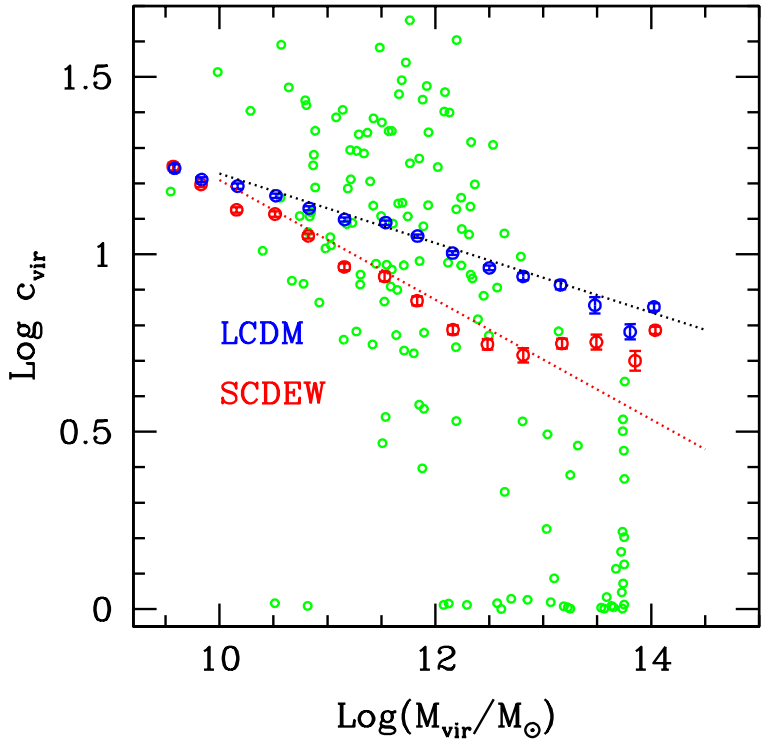}
\end{center}
\vskip -.5truecm
\caption{Halo concentrations predicted by SCDEW and LCDM N--body
  simulations vs.~recent observational outputs, from the SPARC sample
  (courtesy of Katz et al.  \cite{katz}). The large observational
  errors of data, provided in \cite{katz}, are omitted. The visual
  impression that the LCDM fitting line is rather high and not
  sufficiently steep is confirmed by a more quantitative data analysis
  (see text). }
\label{figure2}
\vskip -.3truecm
\end{figure}

This process can be interrupted only if other cosmic components are
involved, before the dissolution is total. SCDEW models therefore
predict a minimal mass scale, below which fluctuations should not
survive the entry in the horizon. The scale below which fluctuations
do not survive this process is 
much lower than the {\it free streaming} mass scale for $\sim 100\,
$eV particles in LWDM models, {ranging around typical Pop III stellar
  mass scales.}

This is why models involving such light particles can form adequate
cosmic structures, as also shown by \cite{MMPB} through N--body
simulations, performed for a model with $m_w=90\, $eV and $\beta=10\,
,$ assuming an {\it ad--hoc} complete coupling cut--off at $z \simeq
50~.$ As expected, DM being made of low mass particles significantly
eases long standing problems. In particular: (i) simulated dwarf
galaxy cores are much flatter than in LCDM; (ii) the number of
MW--size galaxy satellites is reduced by a factor 0.3--0.5, fully
adequate to fit their observed ``scarsity''; (iii) the concentration
distribution is also notably different. Here it may be worth peforming
a comparison between the predicted concentration distribution, already
shown by \cite{MMPB}, and concentrations deduced by \cite{katz} from
the SPARC dataset. This can be attempted on the basis of Figure
\ref{figure2} which seems to favor SCDEW, which provides a closer
approximation to the rapid concentration decrease with mass.

In the Figure, observational point error bars, provided by \cite{katz},
were omitted. The data spread, apparently much larger than
simulations, could however be partially due to observational
uncertainties. Virial mass values are derived from observational
velocities, by assuming a NFW profile and a virial density contrast
$\Delta_v = 98.7$, as in simulations. The black and red dotted curves
are fits of halo concentrations found in LCDM and SCDEW simulations,
respectively. The former one meets all simulation points with an
average quadratic deviation $\sim 10^{-2}$; the deviation for the
latter one is $\sim 4$ times greater, essentially because of the 4 top
mass points, which however include 75 over 4280 simulated halos
(1.75$\, \%$). A requirement for a fair model is that the average
quadratic deviations of observational points from simulation fitting
lines, taken separately for points above and below the fits, are
similar. Such ratio turns out to be 0.385 for LCDM and 1.000 for
SCDEW; the latter value being even unexpectedly close to unity. This
however confirms the visual impression that the LCDM fitting line is
rather high and not sufficiently steep, in respect to data. From these
computations, points with $\log c_{vir} < 0.01$ were omitted.
We plan to perform a more detailed comparison in further work, also to
test whether this kind of data provide any constraints on SCDEW model
parameters.

We conclude that { suitable} SCDEW cosmologies, overlapping LCDM
predictions for scales $\gtrsim 10^{12} h^{-2} M_\odot$
{ can be expected to be quite close to them} down to galactic
scales. They therefore provide an excellent fit of CMB and fluctuation
spectral data. The point being that the new parameters of SCDEW can be
suitably used to improve data fitting.  At still smaller scales we can
however outline a parameter independent SCDEW prediction, { that
  small galaxies form earlier in SCDEW than in LCDM cosmologies,
and that there ought to be a primeval spectrum low--scale cutoff
around the mass scale of Pop III stars.}

\acknowledgments
Thanks are due to Harley Katz and
collaborators for providing us the concentration data files. Andrea
Macci\`o is to be thanked for useful discussions, namely on this very
point. We also acknowledge Sergio Monai's help in the graphic
treatment of some datasets.


\begin{thebibliography}{99}

\bibitem{BM2} S. A. Bonometto, R. Mainini, {\it Growth and dissolution of 
spherical density enhancements in SCDEW cosmologies},  	JCAP06(2017)010,
arXiv:1703.05141



\bibitem{GRV} L.~Amendola, 1999, \prd 60, 043501; T.~Chiba,
  1999, \prd 60, 083508; N.~Bartolo \& M.~Pietroni, 2000, \prd 61,
  023518; F.~Perrotta, C.~Baccigalupi \& S.~Matarrese, \prd 61, 023607;
  G.~Esposito \& D.~Polarski, 2001, \prd 63, 063504; S.M.~Carrol,
  A..~DeFelice, V. Duvvuri, D.A.~Easson, M.~Trodden \& M.S.~Turner,
  2006, \jcap ~0608, 005; L.~Amendola, R.~Gannouji, D.~Polarski \&
  S.~Tsujikawa, 2007, \prd 75, 083504; W.~Hu \& I.~Sawicki, 2007, \prd
  76, 064004; A.~Starobinsky, JEPT Lett 86, 157; S.A. Appleby \&
  R.A.~Battye, 2007, Phys.Lett. B 654, 7; V.~Miranda, S.E.~Joras, I.~Waga
  \& M.~Quartin, 2009, \prl ~102, 2211;  L. Amendola \&
  S.~Tsujikawa, 2010, Dark Energy, Cambridge University Press.

\bibitem{GRM} L.Randall \& R. Sundrum, 1999, \prl ~83,
  3370 \& 4690; P.~Binetruy, C.~Defayet \& Langlois, 2000, \nphysb
  ~565, 269; P.~Binetruy, C.~Defayet, U.~Ellwanger \& Langlois, 2000,
  Phys.Lett B 477, 285; G.R.~Dvali, G.~Gabadadze \& M.~Porrati, 2000,
  Phys.Lett. B 485, 208; C.~Deffayet, G.R.~Dvaali \& G.~Gabadadze,
  2002, \prd 65, 044023; C. Rovelli, Quantum Gravity, 2004,
  Cambridge University Press

\bibitem{DEP}  Ellis J., S. Kalara, K.A. Olive \&
  C. Wetterich, 1989, Phys.Lett. B 228, 264; Ratra B. \& Peebles P.J.E.,
  1988, \prd 37, 3406; Wetterich C., 1995, \aap  ~301, 321;
  L. Amendola, 2010, Dark Energy, Cambridge University Press.

\bibitem{amendola} Amendola L., 1999, \prd  60, 043501;
  Amendola L., 2000, \prd 62, 043511; Amendola L.,
  Tocchini-Valentini D., 2002 \prd 66, 043528;
 
\bibitem{macciobaldi} A.V. Macci\'o, C. Quercellini, R. Mainini,
  L. Amendola \& S.A. Bonometto, 2004,
  \prd 69, 123516, arXiv:astro-ph/0309671; M.~Baldi, V.~Pettorino, G.~Robbers \&
  V.~Springel, 2010, \mnras  ~403, 1684B


\bibitem{BSLV} Silvio A. Bonometto, Giandomenico Sassi, Giu\-sep\-pe
  La Vacca, {\it Dark
  energy from dark radiation in strongly coupled cosmologies with no
  fine tuning}, 2012, \jcap ~08, 015,  arXiv:1206.2281

\bibitem{BM} Silvio A. Bonometto, Roberto Mainini, {\it Fluctuations in strongly coupled cosmologies} 2014, 
 \jcap ~03, 38, arXiv:1311.637

\bibitem{BMM} Silvio A. Bonometto, Roberto Mainini, Andrea
  V. Macci\'o, {\it Stron\-gly Coupled Dark Energy Cosmologies: preserving LCDM success
 and easing low scale problems I - Linear theory revisited}, 2015 \mnras ~453, 1002, arXiv:1503.07875 

\bibitem{MMPB} Andrea V. Macci\'o, Roberto Mainini, Camilla Penzo,
Silvio A. Bonometto, {\it Stron\-gly Coupled Dark Energy Cosmologies: preserving LCDM success and easing low scale problems II - Cosmological
simulations}, 2015, \mnras ~453, 1371, arXiv:1503.07867 


\bibitem{pallosobra} Pallottini A., Ferrara A., Gallerani
  S., Salvadori S., D'Odorico V., 2014, \mnras ~440, 2498; Pallottini
  A., et al., 2015, \mnras ~453, 2465; Sobral D., et al., 2015, ApJ 808,
  139.

\bibitem{Xia}  Xia J.Q., 2013, \jcap ~11, 22

\bibitem{das2006} S.~Das, P.S.~Corasaniti \& J.~Khouri, 2006, \prd 73,
  083509

\bibitem{petre}  P. Petreszky, 2013, Proceedings of the X
  conference on Quark Confinement and Hadron Spectrum, October 2012,
  arXiv:1301.6188v1

\bibitem{universe} Silvio A. Bonometto \& Roberto Mainini, 2016,
Universe 4, 32, arXiv:1610.05519 


\bibitem{SNIA} J.P.~Bernstein, R.~Kessler, S.~Kuhlmann,
  et al., 2012, \apj ~753, 152

\bibitem{demianski} M.~Demianski, E.~Piedipalumbo, D.~Sawant \&
  L.~Amati, 2016,  \aap ~in press, arXiv:1609.09631;
M. Demianski, E. Piedipalumbo, D. Sawant, L. Amati
  arXiv:1610.00854

\bibitem{lensing} R. Laureijs, {\it Euclid
  Definition Study Report}, 2011, arXiv:1110.3193 

\bibitem{MQMAB} A.V.~Macci\'o, C.~Quercellini, R.~Mainini,
  L.~Amendola, S.A.~Bonometto, 2004, \prd 69, 123516, and
  arXiv:astro-ph/0309671

\bibitem{cmbfast} U.~Seljak \& M.~Zaldarriaga, 1996, \apj ~469, 437;
  M.~Zaldarriaga, U.~Seljak \& E.~Bertschinger, 1998, \apj ~494, 491;
  M.~Zaldarriaga \& U.~Seljak, 2000, \apjs ~129, 431.


\bibitem{planck} Planck collaboration, 2015, arXiv:1502.01589v3.


\bibitem{cole} S. Cole et al., 2005, \mnras ~362, 505, arXiv:astro-ph/0501174 

\bibitem{zaroubi} S. Zaroubi, M. Viel, A. Nusser, M. Haehnelt \& T.-S
  Kim, 2006,  \mnras ~369, 734, arXiv:astro--ph/0509563v3


\bibitem{pacucci}  Pacucci F., Pallottini A., Ferrara A. \&
  Gallerani S., 2017, \mnras ~Lett. (in press),  arXiv:1702.04351




\bibitem{katz} Katz H., Lelli F., McGauch S.S., Di Cintio A., Brook
  C.B. Schombert J.M.,  {\it Testing Feedback--Modified Dark Matter Haloes with Galaxy Rotation Curves: Estimation of Halo Parameters and Consistency with LCDM},  2016, arXiv:1605.05971



\end{thebibliography}
\end{document}